\documentclass[12pt]{iopart} 
\usepackage[pdftex]{color,graphicx,hyperref}
\usepackage{bm}
\usepackage{verbatim}
\usepackage{mathrsfs}
\usepackage{amssymb}
\usepackage{exscale}
\newcommand{\beq}{\begin{equation}}
\newcommand{\eeq}{\end{equation}}
\newcommand{\re}[1]{\mbox{Re($  #1$)}}
\newcommand{\im}[1]{\mbox{Im($  #1$)}}

\newcommand{\text}[1]{\mbox{\footnotesize #1}}
\newcommand{\eqref}[1]{(\ref{#1})}
\newcommand{\email}[1]{\begin{indented}
   						\item[]Email: \texttt{\footnotesize\raggedright#1}
   						\end{indented}}

\begin{document}
\title{The Functional Integral formulation of the Schrieffer-Wolff transformation}
\author{Farzaneh Zamani}\email{farzaneh@pks.mpg.de}
\address{Max Planck Institute for the Physics of Complex Systems, 01187 Dresden, Germany}
\address{Max Planck Institute for Chemical Physics of Solids, 01187 Dresden, Germany}
\author{Pedro Ribeiro}\email{ribeiro.pedro@gmail.com}
\address{CeFEMA, Instituto Superior Técnico, Universidade de Lisboa, Av. Rovisco Pais, 1049-001 Lisboa, Portugal}
\author{Stefan Kirchner}\email{stefan.kirchner@correlated-matter.com}
\address{Center for Correlated Matter, Zhejiang University, Hangzhou,  Zhejiang 310058, China}
\address{Max Planck Institute for the Physics of Complex Systems, 01187 Dresden, Germany}

\date{today}
\begin{abstract}
We revisit the Schrieffer-Wolff transformation and present a path integral version of this important canonical transformation. 
The equivalence between the low-energy sector of the Anderson model in the so-called {local moment} regime and the spin-isotropic Kondo model is usually established via a canonical transformation performed on the Hamiltonian, followed by a projection.
Here we present a path integral {formulation} of the Schrieffer-Wolff transformation which relates  the functional integral form of the partition function of the Anderson model to that of its effective low-energy model. The resulting functional integral assumes the form of  a spin path integral  and includes a geometric phase factor, {\it i.e.} a Berry phase.
Our approach stresses the underlying symmetries of the model and allows for a straightforward generalization of the transformation to more involved models. It thus not only sheds new light on a classic problem, it also offers a systematic route of obtaining effective low-energy models and higher order corrections.
This is demonstrated by obtaining the effective low-energy model of a quantum dot attached to two ferromagnetic leads. 
\end{abstract}


\maketitle
\section{Introduction and Motivation}
\label{intro}
Canonical transformations have played a key role in the development of various branches of physics. They  remain an important tool in tackling new problems, often capturing the {essence of the underlying physics}. This is {\it e.g,} the case for the Firsov-Lang transformation applied to the polaron problem or the Bogolibov transformation of superconductivity~\cite{Firsov.68,Hohenadler.07,Bogoliubov.58,Valatin.58}.\\
A particular important transformation is the  
Schrieffer-Wolff  transformation~\cite{Schrieffer.66}, {\it i.e.}, 
a canonical transformation applied to a Hamiltonian $H$ which is followed by a projection into a subspace of the Hilbert space associated with $H$ with the aim of obtaining an effective Hamiltonian for the low-energy sector of  $H$. The effective Hamiltonian is usually obtained perturbatively and the calculation of higher-order terms is often cumbersome. 
The Schrieffer-Wolff transformation was originally introduced by Schrieffer and Wolff to demonstrate that the low-energy behavior of the 
Anderson model in the local-moment regime is that of a quantum spin coupled isotropically via an anti-ferromagnetic exchange interaction  to the local spin-density of an otherwise free conduction band~\cite{Schrieffer.66}. This is accomplished by constructing the generator  of a canonical transformation which removes charge fluctuations from the effective Hamiltonian in lowest order in the hybridization, followed by a projection into the singly occupied subspace of the local (impurity) Hilbert space.
An alternate form of the Schrieffer-Wolff transformation from the Anderson to the Kondo model is due to Hewson~\cite{Hewson}. In Ref.~\cite{Hewson}, the effective Hamiltonian in the singly occupied subspace is constructed by starting from the Schr\"{o}dinger equation  and eliminating the components of the groundstate wavefunction in the empty and doubly occupied subspace.
In recent years, there has been renewed interest in the Schrieffer-Wolff transformation~~\cite{Kehrein.96,Chana.04,Thomas.09,Bravyi.11,Ong.11,Kessler.12}. 
A detailed review and a compilation of rigorous results of the Schrieffer-Wolff transformation  can be found in Ref.~\cite{Bravyi.11}. The application of the Schrieffer-Wolff method to systems coupled to dissipative environments  appeared in Ref.~\cite{Kessler.12} while Refs.~\cite{Thomas.09,Ong.11} reported applications to systems that contain more than one quantum impurity.

One of the major difficulties with the Schrieffer-Wolff transformation in either of these operator based versions is the determination of
higher order terms beyond those quadratic in the hybridization between the local and conduction electrons.  This makes generalizations to more complex models tedious.
Here, we will present a path integral formulation of the Schrieffer-Wolff transformation which not only simplifies the construction of higher order terms of the  transformation as the operator algebra is replaced by (anti-)commuting fields but which can also  be straightforwardly generalized to more complex situation like {\it e.g.} interacting bath modes. Moreover, our approach brings out the geometric or Berry phase associated with dynamics in the reduced Hilbert space~\cite{Inomata,Schulman} and allows for an analysis of the effect of charge fluctuations on the Berry phase term. 
Among the possible applications {of our approach} are multi-impurity systems and systems with generalized baths. This may not only be of relevance in addressing the effect of charge fluctuations in Kondo lattice systems. It should generally prove useful {whenever} the topological term generated by restricting the dynamics to the sub-space turns out to be  non-trivial~\cite{Jizba}.
A better understanding of Berry phase effects may also shed new light on {certain} quantum phase transitions where dynamics is {an integral} part of criticality~\cite{Kirchner.10} and where the Berry phase term in the associated effective action
invalidates a naive quantum-to-classical mapping~\cite{Kirchner.08}.
Last but not least, our approach might prove useful in constructing effective models for the real-time dynamics of nano-electro-mechanical systems~\cite{Koenig.12}.
 
The outline of the paper is as follows. In Section \ref{model} we introduce the Anderson and Kondo model and briefly review the traditional  {operator-based} formulation of the Schrieffer-Wolff transformation. Section \ref{PIversion} introduces  our path integral version of the Schrieffer-Wolff transformation. In Section \ref{Sec:gauge} two explicit forms of the dynamical or Berry phase term of the spin path integral are obtained depending on the explicit form of the gauge transformation performed in Section \ref{PIversion}. Section \ref{Sec:gauge} ends with a discussion of the effect of different Hubbard-Stratonovich transformations on the Schrieffer-Wolff transformation. 
In Section \ref{Sec:magT}, and we apply the path integral version of the Schrieffer-Wolff transformation to the magnetic single-electron transistor, where the leads are formed by ferromagnetic metals. As expected~\cite{Kirchner.05a} the effective low-energy model is a sub-Ohmic Bose-Fermi Kondo model~\cite{Zhu.02}. In \ref{AppA}, we derive the effective action of a metallic ferromagnet which is needed in Section \ref{Sec:magT}. The concluding section, {\it i.e.} Section \ref{Sec:conclusion},  contains a brief summary of our results.
 
%

\section{The Anderson and Kondo model}
\label{model}
The Anderson impurity model is a paradigmatic model of strong electron correlations. It describes an impurity state that can at most be doubly occupied by electrons with spin quantum number $\sigma=\pm$ and which hybridizes with conduction electrons of an otherwise uncorrelated electron band. It is defined by the Hamiltonian
\begin{equation}
\label{eq:Anderson}
\fl
H_{A}= \sum_{\sigma=\pm} \epsilon_{d}^{} d_{\sigma}^{\dagger}d_{\sigma}^{} + U d_{+}^{\dagger}d_{+}^{} d_{-}^{\dagger}d_{-}^{}
+ \sum_{{\bf k},\sigma=\pm}\big ( V_{\bf k} c^\dagger_{{\bf k}\sigma}d^{}_{\sigma} + V_{\bf k}^{*} d^{\dagger}_{\sigma}c^{}_{{\bf k}\sigma} \big) + \sum_{{\bf k},\sigma=\pm} \epsilon_{k}^{} c^\dagger_{{\bf k}\sigma} c^{}_{{\bf k}\sigma},
\end{equation}
where $\epsilon_d$ is the energy of the singly occupied impurity state with respect to the Fermi energy of the conduction band, $U$ is the Coulomb integral, and $V_{{\bf k}}$ is a measure of the strength of hybridization between the local and conduction electron states. If $\epsilon_d<0$, $\epsilon_d+U>0$, with $|\epsilon_d|\gg \Delta(0)$, and   $|\epsilon_d+U|\gg \Delta(0)$, where $\Delta(\epsilon)=\pi\sum_{\bf k}|V_{{\bf k}}|^2 \delta(\epsilon-\epsilon_k)$, the low-energy sector of the Anderson model is equivalent to the Kondo model plus
a potential scattering term
\begin{eqnarray}
\fl
H_{K}&=&  \sum_{{\bf k},\sigma=\pm} \epsilon_{k}^{} c^\dagger_{{\bf k}\sigma} c^{}_{{\bf k}\sigma} \,+\, \frac{1}{2} 
\sum_{\stackrel{{\bf k},{\bf k \prime}}{\sigma,\sigma \prime}}
J_{{\bf k},{\bf k \prime}} {\bf S}\cdot  c^{\dagger}_{{\bf k}\sigma} {\mathbf \sigma }c^{}_{{\bf k\prime}\sigma \prime} +\sum_{{\bf k},{\bf k \prime}} W^{\mbox{\tiny pot}}_{{\bf k},{\bf k \prime}} c^\dagger_{{\bf k \prime}\sigma} c^{}_{{\bf k}\sigma}.
\end{eqnarray}	
This equivalence in the low-energy sector is established through a canonical transformation~\cite{Schrieffer.66}
\begin{equation}
\label{eq:generator}
\tilde{H}=e^S H_{A} e^{-S},
\end{equation}
where hermiticity of the Hamiltonian implies $S^\dagger=-S$ and the generator $S$ of the transformation is chosen such that the hybridization vanishes {to} lowest order in $V_{\bf k}$,
\begin{equation}
\label{SWT}
\Big [ H_{A}-\sum_{{\bf k},\sigma=\pm}\big ( V_{\bf k} c^\dagger_{{\bf k}\sigma}d^{}_{\sigma} + V_{\bf k}^{*} d^{\dagger}_{\sigma}c^{}_{{\bf k}\sigma} \big) ,S \Big ]=\sum_{{\bf k},\sigma=\pm}\big ( V_{\bf k} c^\dagger_{{\bf k}\sigma}d^{}_{\sigma} + V_{\bf k}^{*} d^{\dagger}_{\sigma}c^{}_{{\bf k}\sigma} \big),
\end{equation}
followed by a projection into the singly occupied subspace. 
Eq.~(\ref{SWT}) can formally be solved by introducing Liouville operators ${\mathcal{L}_{H_x}}$ defined by
\begin{equation}
\label{eq:Liouville}
{\mathcal{L}_{H_{x}}}A=\big[H_x,A \big ].
\end{equation}
As a result, one finds that
up to second order in the hybridization $J_{{\bf k},{\bf k \prime}}$ and $W^{\mbox{\tiny pot}}_{{\bf k},{\bf k \prime}}$ are given by
\begin{equation}
J_{{\bf k},{\bf k \prime}}= V_{\bf k}  V_{\bf k\prime}^{*}\Big\{ \frac{2 U}{(U+\epsilon_d)\epsilon_d} \Big\}~\mbox{and}~ 
W^{\mbox{\tiny pot}}_{{\bf k},{\bf k \prime}}= \frac{V_{\bf k}  V_{\bf k\prime}^{*}}{2}\Big\{ \frac{1}{U+\epsilon_d} +\frac{1}{\epsilon_d}\Big\}. 
\end{equation}	
Schrieffer and Wolff derived additional terms that can safely be neglected in the standard Kondo case but may become important, {\it} if the impurity is immersed  in a superconductor~\cite{Schrieffer.66}.

\section{The path integral version of the Schrieffer-Wolff transformation}
\label{PIversion}
In this section we will derive the path integral version of the Schrieffer-Wolff transformation.
In its most general form, a Schrieffer-Wolff transformation creates an effective Hamiltonian describing the dynamics in a sub space of the total phase space of the original problem by reducing the number of Fock states accessible to the system.
A path integral reformulation of this type of canonical transformations is highly desirable as
the functional integral description  avoids the cumbersome
anti- or commutator algebra and often offers insights into the problem that are hard to obtain via the operator calculus. This may be particularly
relevant when a knowledge of higher order corrections is required. 
The path integral is the tool of choice when it comes to integrating out 
part of the fluctuation spectrum, {\it e.g.} charge fluctuations in the case of the standard Kondo problem, or to implement holonomic constraints
on the dynamics of the system. 
A restriction to a sub space of the original Fock space will in general lead to a  geometric phase term that {reflects the dynamics restricted to the sub-space}. In order to identify the resulting action as that associated with an effective Hamiltonian, both the Hamiltonian part  and
the geometric part of the action have to be local in (imaginary) time. This requires {an} identification of the underlying manifold as a group manifold and points to the importance of properly treating the symmetries inherent to the problem.
Naturally, one expects that a path integral version exists for a canonical transformation, like the Schrieffer-Wolff transformation, that can be performed on the Hamiltonian level.
One property that should come out of a proper path integral treatment of the Schrieffer-Wolff transformation is the 
spin-isotropy of the effective model associated with the  low-energy sector of the Anderson model. 

At this point it is useful to realize that the interaction term of the local part of  the Anderson Hamiltonian  possesses spin-rotational invariance. This can {\it e.g.} be seen by noticing that Eq.~(\ref{eq:Anderson}) is equivalent to
\begin{eqnarray}
\label{eq:Anderson-v2}
\fl
H_{A}&=& \sum_{\sigma=\pm} E_{d}^{} d_{\sigma}^{\dagger}d_{\sigma}^{} - \frac{2U}{3} {\boldsymbol{S}^2}
+ \sum_{{\bf k},\sigma=\pm}\big ( V_{\bf k} c^\dagger_{{\bf k}\sigma}d^{}_{\sigma} + V_{\bf k}^{*} d^{\dagger}_{\sigma}c^{}_{{\bf k}\sigma} \big) + \sum_{{\bf k},\sigma=\pm} \epsilon_{k}^{} c^\dagger_{{\bf k}\sigma} c^{}_{{\bf k}\sigma},
\end{eqnarray}
where $\boldsymbol{S}=\frac{1}{2}\sum_{\alpha,\beta} d^\dagger_{\alpha}\boldsymbol{\sigma}_{\alpha \beta} d^{}_{\beta}$ and $E_{d}^{}=\epsilon_{d}^{}+U/2$. 
It thus is necessary to perform the transformation without artificially reducing the invariances of the action, {\it i.e.} breaking spin-rotational invariance~\cite{Schulz.90}. 
We  start from 
the partition function of the Anderson model, Eq.~(\ref{eq:Anderson-v2}), in terms of a functional integral~\cite{Hamann.70}
\begin{eqnarray}
\label{Andersonpartition}
Z=\int {\mathcal{D}}[\bar{\psi},\psi, \bar{\phi},\phi] e^{-S[\bar{\psi},\psi, \bar{\phi},\phi]},
\end{eqnarray}
where the action $S=\int_0^\beta~ L~d\tau$ is given by
\begin{eqnarray}
\label{AndersonAction}
\fl
S[\bar{\psi},\psi, \bar{\phi},\phi]=\int_0^\beta d\tau \Bigg\{ \sum_{\sigma=\pm}  \bar{\psi}_\sigma(\tau) [\partial_{\tau}+\epsilon_d] \psi_\sigma(\tau)
+U  \bar{\psi}_+(\tau)\bar{\psi}_-(\tau)\psi_-(\tau)\psi_+(\tau)\Big . \nonumber\\
\fl +\sum_{{\mathbf{k}},\sigma=\pm}\big [ V^{}_{k} \bar{\psi}_\sigma(\tau) \phi_\sigma({\mathbf{k}},\tau) +V^{*}_{k} \bar{\phi}_\sigma({\mathbf{k}},\tau)\psi_\sigma(\tau) \big ]
+ \Big .  \sum_{{\mathbf{k}},\sigma} \bar{\phi}_\sigma({\mathbf{k}},\tau) [\partial_\tau +\epsilon_k-\mu] \phi_\sigma({\mathbf{k}},\tau) \Bigg \},
\end{eqnarray}
and where $\bar{\psi}_\sigma(\tau),\psi_\sigma(\tau), \bar{\phi}_\sigma({\mathbf{k}},\tau)$, and $\phi_\sigma({\mathbf{k}},\tau)$ are Grassmann fields related to the  $d^{\dagger}_{\sigma}$, $d^{}_{\sigma}$, $c^\dagger_{{\bf k}\sigma}$ and $c^{}_{{\bf k}\sigma}$ operators. The explicit imaginary time ($\tau$) dependence of the fields will be suppressed in what follows.  
As our goal is to integrate out charge fluctuations, it is useful to recast the  quartic term into a charge and a spin part
\begin{equation}
\label{eq:decoupling_v1}
U n_+n_-=\frac{U}{4}n^2_{}-U S_z^2,
\end{equation}
with $n=n_++n_-$ and $S^z=(n_+-n_-)/2$,
which allows for a Hubbard-Stratonovich decoupling of the action via
\begin{eqnarray}
\fl
\exp{[-U n_+n_-]}=\frac{1}{\pi U} \int d\Delta \int dm \exp{[-\frac{1}{U}(\Delta^2+m^2)+i\Delta n+2m S^z]}
\end{eqnarray}
and the help of two bosonic decoupling fields $\Delta$ and $m$.
The partition function can then be written as
\begin{eqnarray}
\label{eq:decoupled}
Z=\int {\mathcal{D}}[\Delta]\int {\mathcal{D}}[m]\int {\mathcal{D}}[\bar{\psi},\psi, \bar{\phi},\phi]\,e^{-S[\bar{\psi},\psi, \bar{\phi},\phi,\Delta,m]},\nonumber \\
\fl S[\bar{\psi},\psi, \bar{\phi},\phi,\Delta,m]=
\int_0^\beta d\tau \Bigg\{  \boldsymbol{\bar{\psi}} (\partial_{\tau}+\epsilon_d) \boldsymbol{\psi}+\frac{1}{U}(\Delta^2+m^2)-i\Delta \boldsymbol{\bar{\psi}}\boldsymbol{{\psi}}-m \boldsymbol{\bar{\psi}}\sigma^z_{}\boldsymbol{{\psi}}
\Big . \nonumber\\
 +\sum_{{\mathbf{k}}}\big [ V^{}_{k} \boldsymbol{\bar{\psi}} \boldsymbol{\phi}({\mathbf{k}}) +V^{*}_{k} \boldsymbol{\bar{\phi}}({\mathbf{k}})\boldsymbol{\psi} \big ]
- \Big .  \sum_{{\mathbf{k}}} \boldsymbol{\bar{\phi}}({\mathbf{k}})\mathcal{G}^{-1}_c(\tau,k) \boldsymbol{\phi}({\mathbf{k}}) \Bigg \},
\end{eqnarray}
where we introduced $\mathcal{G}_c(\tau,k)=[-\partial_\tau -\epsilon_k +\mu]^{-1}$ and the spinor notation $\boldsymbol{\bar{\psi}}=(\bar{\psi}_+\, \bar{\psi}_-)$, so that $\boldsymbol{\bar{\psi}} \boldsymbol{\psi}=\bar{\psi}_+{\psi}_++\bar{\psi}_-{\psi}_-$, $\boldsymbol{\bar{\psi}}\sigma^z_{}\boldsymbol{{\psi}}=2S^z$ and likewise for $\boldsymbol{\bar{\phi}}({\mathbf{k}})$.
The decoupled action 
breaks at least formally the underlying spin-rotational invariance of Eq.~(\ref{AndersonAction}) which leads to incorrect excitation spectra near saddle point solutions of Eq.~(\ref{eq:decoupled})~\cite{Hamann.70,Schulz.90} and a proper incorporation of fluctuations around these saddle points is vital to restore spin-rotational invariance.
A general solution to this problem was discussed by Schulz in Ref.~\cite{Schulz.90} which we will follow here.  To this end we note that the choice of
spin quantization axis is arbitrary. We can exploit that the effective action remains invariant under a rotation of the quantization axis, $\sigma^3 \longrightarrow \boldsymbol{\Omega}\cdot \boldsymbol{\sigma}$, by  summing over all possible choices $\boldsymbol{\Omega}$, properly normalized, to ensure a rotationally invariant saddle point~\cite{Schulz.90,Kopec.13}
\begin{eqnarray}
\label{eq:action-rotinvariant}
Z=\int {\mathcal{D}}[\boldsymbol{\Omega}] Z[\Omega].
\end{eqnarray} 
Next, a unitary transformation on the local Grassmann fields is performed, {\it i.e.} $\boldsymbol{\bar{\chi}}=\boldsymbol{\bar{\psi}} U$, $\boldsymbol{\chi}=U^\dagger\boldsymbol{\psi} $ which leaves the measure invariant, ${\mathcal{D}}[\boldsymbol{\hat{\psi}}\boldsymbol{\psi}]={\mathcal{D}}[\boldsymbol{\hat{\chi}}\boldsymbol{\chi}]$.

We thus arrive at
\begin{eqnarray}
Z=\int {\mathcal{D}}[\boldsymbol{\Omega}]\int{\mathcal{D}}[\boldsymbol{\bar{\phi}},\boldsymbol{\phi}]
\int {\mathcal{D}}[\Delta,m,\boldsymbol{\hat{\chi}},\boldsymbol{\chi}]\,
e^{-S_{\mbox{\tiny eff}}[\boldsymbol{\bar{\chi}},\boldsymbol{\chi}, \boldsymbol{\bar{\phi}},\boldsymbol{\phi},\Delta,m]},\nonumber \\
\fl S_{\mbox{\tiny eff}}[\boldsymbol{\bar{\chi}},\boldsymbol{\chi}, \boldsymbol{\bar{\phi}},\boldsymbol{\phi},\Delta,m]=
\int_0^\beta d\tau \Bigg\{  \boldsymbol{\bar{\chi}} (\partial_{\tau}+\epsilon_d+U^\dagger \partial_{\tau}U) \boldsymbol{\chi}+\frac{1}{U}(\Delta^2+m^2) \Big . \nonumber\\
\fl - \sum_{{\mathbf{k}}} \boldsymbol{\bar{\phi}}({\mathbf{k}})\mathcal{G}^{-1}_c(\tau,k) \boldsymbol{\phi}({\mathbf{k}})
 - i\Delta \boldsymbol{\bar{\chi}}\boldsymbol{{\chi}}-m \boldsymbol{\bar{\chi}}\sigma^z_{}\boldsymbol{{\chi}}
+\sum_{{\mathbf{k}}}\big [ V^{}_{k} \boldsymbol{\bar{\chi}} \boldsymbol{U}\boldsymbol{\phi}({\mathbf{k}}) +V^{*}_{k} \boldsymbol{\bar{\phi}}({\mathbf{k}})\boldsymbol{U}^{\dagger}\boldsymbol{\chi} \big ]
 \Big .  \Bigg \}.
 \label{eq:decoupledAction}
\end{eqnarray}
As the preceding expression is at most quadratic in the local Grassmann fields, integrating $\boldsymbol{\psi}$ out, yields
\begin{eqnarray}
\label{decoupledActionII}
%
\fl
S_{\mbox{\tiny eff}}[\boldsymbol{\bar{\phi}},\boldsymbol{\phi},\Delta,m]
=
\int_0^\beta d\tau \Big\{\frac{1}{U}(\Delta^2+m^2)\Big . \Big.  \nonumber \\
\fl
+\sum_{{\mathbf{k}},{\mathbf{k'}}}\big (  \boldsymbol{\bar{\phi}}({\mathbf{k}}) V^{*}_{k} \boldsymbol{U}^{\dagger}G_{d}\boldsymbol{U}  V^{}_{k'} \boldsymbol{\phi}({\mathbf{k'}}) \big)
-\sum_{{\mathbf{k}}} \boldsymbol{\bar{\phi}}({\mathbf{k}}) \mathcal{G}^{-1}_c \boldsymbol{\phi}({\mathbf{k}}) \Big \} 
- \Tr \ln \Big[  -G_d^{-1}\Big],
\end{eqnarray}
where we introduced the local Green function 
$G_d^{-1}={\mathcal{G}}_d^{-1}-\Sigma$ with ${\mathcal{G}}_d^{-1}=-(\partial_\tau+\epsilon_d-i\Delta_0)+ m_0\sigma^3$ and $\Sigma= \boldsymbol{U}\partial_\tau \boldsymbol{U}^\dagger+\delta m(\tau)\sigma^3+i\delta \Delta(\tau)$ and we have split the fields $\Delta(\tau)$ and $m(\tau)$ into their static, {\it i.e.} $\Delta_0$ and $m_0$,   and $\tau$-dependent parts.

So far, our treatment has been exact and Eq.(\ref{decoupledActionII}) is a faithful representation of the action associated with the Anderson model, Eq.(\ref{eq:Anderson}).
Under the assumption that the terms contributing to $\Sigma$ are small compared to ${\mathcal{G}}_d^{-1}$ , {\it i.e.} $|\boldsymbol{U}\partial_\tau \boldsymbol{U}^\dagger|$, 
$|\delta \Delta(\tau)|$, $|\delta m(\tau)|\ll|\epsilon_d-i\Delta_0-m_0 \sigma^3|$, we can approximate $Tr \ln (- G_d^{-1})$ by
\begin{equation}
\Tr \ln \Big[- G_d^{-1}\Big]=\Tr \ln \Big[- \mathcal{G}_d^{-1}\Big] - \Tr  \Big[\mathcal{G}_d \Sigma \Big] -  \ldots
\end{equation}
As we are interested in obtaining an effective low-energy limit of  Eq.~(\ref{eq:Anderson-v2}) in the Kondo regime, where $\epsilon_d<0,~U>0$ and $-|V|^2/\epsilon_d, |V|^2/|U|\ll 1$, we  will  ignore fluctuations around the static charge configuration  so that $\Sigma \approx \boldsymbol{U}\partial_\tau \boldsymbol{U}^\dagger$.
The saddle point values $\Delta_0$ and $m_0$ are obtained from $\partial_{\Delta_0} \ln Z =0$ and   $\partial_{m_0} \ln Z =0$:
\begin{eqnarray}
\label{eq:HS1}
\frac{2}{U}\Delta_0 &=& \frac{\delta}{\delta \Delta} \Tr \ln \Big[- G_d^{-1}\Big]=i,
\end{eqnarray}
where the right hand side holds in the Kondo regime and is equivalent to $\sum_{\sigma=\pm}\langle d^\dagger_{\sigma} d^{}_{\sigma}\rangle=1$.  
Under similar conditions, we find for $m_0$
 \begin{eqnarray}
 \label{eq:HS2}
\frac{2}{U} m_0 &=& \frac{\delta}{\delta m} \Tr \ln \Big[- G_d^{-1}\Big]= \Tr [ G_d \sigma^z ]= 1
\end{eqnarray}
or $m_0=U/2$. Away from $\epsilon_d+U/2=0$, corrections {to $\Delta_0$ and $m_0$} are exponentially small in $1/T$ and will be ignored in what follows.
With these values for $\Delta_0$ and $m_0$, we find
\begin{eqnarray}
\label{eq:couplings}
\fl
\boldsymbol{U}^\dagger G_{d}(\tau)  \boldsymbol{U} \approx  \boldsymbol{U}^\dagger \frac{1}{-\epsilon_d+i\Delta_0+m_0\sigma^z}  \boldsymbol{U}=2 \boldsymbol{U}^\dagger \Big (\frac{ U \sigma^z-(2\epsilon_d+U)}{(2\epsilon_d+U)^2 -U^2} \Big ) \boldsymbol{U}\nonumber \\
\fl
= \frac{U}{|\epsilon_d|(\epsilon_d+U)} \boldsymbol{\Omega}\cdot \frac{\boldsymbol{\sigma}}{2} + \frac{\epsilon_d+U/2}{|\epsilon_d|(\epsilon_d+U)},
\end{eqnarray}
where $\epsilon_d<0$, appropriate for the local moment regime of the Anderson model, was used.
The second term in this expression describes a potential scattering contribution that vanishes for a particle-hole symmetric model, {\it i.e.} $U=-2\epsilon_d$. 
Finally, $Tr  \Big[\mathcal{G}_d \Sigma \Big]$ needs to be analyzed in the local moment regime:
\begin{eqnarray}
\label{eq:Berry}
\fl
-\Tr  \Big[\mathcal{G}_d \Sigma \Big]= \tr \Big\{ \frac{1}{\beta} \sum_{\omega_n} \frac{1}{i\omega_n-\epsilon_d-U/2-U\sigma^z} \int d\tau {\mathbf U}\frac{\partial}{\partial \tau} {\mathbf U}^\dagger \Big \}\nonumber \\
\fl
=\int d\tau  {\mathbf U}\frac{\partial}{\partial \tau} {\mathbf U}^\dagger\Big|_{1,1}-\Big ( \frac{e^{2\beta \epsilon_d}}{e^{\beta(U/2+\epsilon_d)}+e^{2\beta \epsilon_d}} +\frac{e^{-\beta U}}{e^{\beta(U/2+\epsilon_d)}+e^{-\beta U}} \Big ) \int d\tau  {\mathbf U}\frac{\partial}{\partial \tau} {\mathbf U}^\dagger\Big|_{1,1},
\end{eqnarray}
where $\tr$ is the trace in spin space only and we have used that $ {\mathbf U}\frac{\partial}{\partial \tau} {\mathbf U}^\dagger\Big|_{1,1}=- {\mathbf U}\frac{\partial}{\partial \tau} {\mathbf U}^\dagger\Big|_{2,2}$, see Sec.~\ref{Sec:gauge} .
In the Kondo regime, where $\epsilon_d<0$, $U>0$ and $-|V|^2/\epsilon_d, |V|^2/|U|\ll 1$, the second term of the right hand side of Eq.~(\ref{eq:Berry}) is exponentially small. 
In this case
\begin{equation}
-Tr  \Big[\mathcal{G}_d \Sigma \Big]\approx \int d\tau  {\mathbf U}\frac{\partial}{\partial \tau} {\mathbf U}^\dagger\Big|_{1,1} ,
\end{equation}
which is purely imaginary.
Various choices for ${\mathbf U}$ and thus the Berry phase term will be discussed in Sec.~\ref{Sec:gauge}.

\noindent
Collecting all terms, the final form of the partition function of Eq.~(\ref{eq:Anderson}) in the Kondo regime is
\begin{eqnarray}
\label{eq:PIKondo}
Z= \int {\mathcal{D}}[\boldsymbol{\Omega}(\theta,\phi)]\int{\mathcal{D}}[\boldsymbol{\bar{\psi}},\boldsymbol{\psi}]\,e^{-S_{\mbox{\tiny eff}}[\boldsymbol{\Omega}, \boldsymbol{\bar{\psi}},\boldsymbol{\psi}]}, \\
S_{\mbox{\tiny eff}}[\boldsymbol{\Omega}, \boldsymbol{\bar{\psi}},\boldsymbol{\psi}] =
\int d\tau  {\mathbf U}\frac{\partial}{\partial \tau} {\mathbf U}^\dagger\Big|_{1,1} + \nonumber \\ 
\int_0^\beta d\tau \Bigg\{\frac{1}{2}\sum_{{\mathbf{k},\mathbf{k}'}} J_{k,k'} \boldsymbol{\Omega}\cdot \boldsymbol{\bar{\psi}}_{\mathbf{k}} \frac{\boldsymbol{\sigma}}{2}\boldsymbol{\psi}_{\mathbf{k}'} + \sum_{{\mathbf{k},\mathbf{k}'}} W_{k,k'}\boldsymbol{\bar{\psi}}_{\mathbf{k}}\boldsymbol{\psi}_{\mathbf{k}'}
- \sum_{{\mathbf{k}}} \boldsymbol{\bar{\psi}}_{\mathbf{k}}\mathcal{G}^{-1}_c(\tau,k) \boldsymbol{\psi}_{\mathbf{k}}
 \Big .  \Bigg \}, \nonumber
\end{eqnarray}
which is a standard spin path integral representation~\cite{Inomata,Kuratsuji.80} of the spin-isotropic Kondo model based on spin-coherent (and fermionic) coherent states. We showed that starting from the Anderson model in the local moment regime, an effective Kondo model  can be obtained with antiferromagnetic exchange coupling $J_{k,k'}=\frac{2 V^{}_{k} V^{*}_{k'} U}{|\epsilon_d|(\epsilon_d+U)}>0$ and an additional potential scattering term with potential strength $W_{k,k'}=V^{}_{k} V^{*}_{k'}\frac{\epsilon_d+U/2}{|\epsilon_d|(\epsilon_d+U)}$.
The Hamiltonian associated with Eq.~(\ref{eq:PIKondo}) is the Kondo Hamiltonian 
\begin{equation}
\label{eq:KondoHam}
H_{\mbox{\tiny KM}}=J_K{\boldsymbol{S}}\cdot \boldsymbol{s}_c(0)+ \sum_{{\mathbf{k}},\sigma} \epsilon_k c^\dagger_{{\mathbf{k}},\sigma}c^{}_{{\mathbf{k}}\sigma}+\sum_{\sigma}\sum_{{\mathbf{k},\mathbf{k}'}}  W c^\dagger_{{\mathbf{k}},\sigma}c^{}_{{\mathbf{k'}}\sigma}~,
\end{equation}
where the conduction electron spin density $\boldsymbol{s}_c(0)$ at the impurity site is $\boldsymbol{s}_c(0)=\sum_{\sigma,\sigma'}\sum_{{\mathbf{k},\mathbf{k}'}} c^\dagger_{{\mathbf{k}},\sigma} \frac{\boldsymbol{\sigma}}{2}c^{}_{{\mathbf{k'}}\sigma'}$.
The local spin excitations are encoded in the functional integral over the sphere  $S^2=SU(2)/U(1)$ parametrized by $\theta(\tau)$ and $\phi(\tau)$.
It follows from Eq.~(\ref{eq:PIKondo}) that in the case $J_K=0$ the action associated with a free quantum spin ${\boldsymbol{S}}$ is just $\int d\tau  {\mathbf U}\frac{\partial}{\partial \tau} {\mathbf U}^\dagger\Big|_{1,1}$. 
 This term is the analog of $\int d\tau  \overline{\alpha} \partial_{\tau} \alpha$ in the standard path integral of a bosonic field $\alpha$. The underlying finite Hilbert space of the spin problem and the resulting compact group space is reflected in the geometric nature of $\int d\tau  {\mathbf U}\frac{\partial}{\partial \tau} {\mathbf U}^\dagger\Big|_{1,1}$\footnote{An introduction into spin-coherent states and the spin-path integral can {\it e.g.} be found in~\cite{Kuratsuji.80,Perelomov,Inomata,Wen,Fradkin}.}.

It will be shown in Sec.~\ref{Sec:gauge} that  $\int d\tau  {\mathbf U}\frac{\partial}{\partial \tau} {\mathbf U}^\dagger\Big|_{1,1}$ is purely imaginary and is equal to
$i\,\frac{1}{2} \int d\tau (1-\cos{\phi(\tau)})\dot \theta(\tau)$. This is the so-called Berry phase term. It is a geometric phase factor that equals the area traced out on the sphere $S^2$ by each closed path entering the path integral.\\
Our path integral version of the Schrieffer-Wolff transformation not only demonstrates how the spin path integral emerges in the Kondo regime of the Anderson model, it also gives a straightforward tool to evaluate corrections and higher order contributions. The corrections to the Berry phase term away from the Kondo regime {\it e.g.} follow from Eq.~(\ref{eq:Berry}). Note that the presence of a magnetic field, taken along the direction of the local quantization axis, does not invalidate the 
Schrieffer-Wolff transformation as long as the assumptions  entering Eqs.~(\ref{eq:HS1}) and (\ref{eq:HS2}) are fulfilled. 
 
\section{Gauge transformation, stereographic projection, and  Hubbard-Stratonovich decouplings}
\label{Sec:gauge}
In this section we will discuss how different choices for the gauge transformation on the local degree of freedom, see Sec.~\ref{PIversion}, affect the final form for the path integral version in our form of the Schrieffer-Wolff transformation. 
It was shown in Sec.~\ref{PIversion} that the term 
$\int d\tau  {\mathbf U}\frac{\partial}{\partial \tau} {\mathbf U}^\dagger\Big|_{1,1}$ appears in the action associated with the Kondo model and {represents} the dynamic
part of the action of a quantum spin. 
As discussed in  Sec.~\ref{PIversion}, the Berry phase term of the spin path integral 
for the Kondo model originates from the term
\begin{equation}
-\Tr  \Big[\mathcal{G}_d \Sigma \Big]= \tr \Big\{ \frac{1}{\beta} \sum_{\omega_n} \frac{1}{i\omega_n-\epsilon_d-U/2-U\sigma^z} \int d\tau {\mathbf U}\frac{\partial}{\partial \tau} {\mathbf U}^\dagger \Big \},
\end{equation}
which leads us to consider ${\mathbf U}\frac{\partial}{\partial \tau} {\mathbf U}^\dagger$.\\
For a general unitary $n\times n$-dimensional matrix $\mathbf U$ it follows from $\mathbf{U} \mathbf{U}^{\dagger}_{}=1$ that
\[\mbox{Tr}\{\mathbf{U U}^{\dagger}_{}\}=n. \]
This implies 
that $\mbox{Tr}\{\mathbf{U}_{}^{\dagger}\partial_{\tau}\mathbf{U}^{}_{}\}=-[\mbox{Tr}\{ \mathbf{U}_{}^{\dagger}\partial_{\tau}\mathbf{U}^{}_{} \}]^{*}$. In other words, $\sum_{i}\mathbf{U}_{}^{\dagger}\partial_{\tau}\mathbf{U}^{}_{}\big|_{i,i}=-\sum_{i}\big(\mathbf{U}_{}^{\dagger}\partial_{\tau}\mathbf{U}^{}_{}\big|_{i,i}\big)^{*}$
Therefore, we can conclude that the diagonal elements of  $\mathbf{U}_{}^{\dagger}\partial_{\tau}\mathbf{U}^{}_{}$ are purely imaginary, if ${\mathbf U}$ is a unitary matrix. 

In the present case,
${\mathbf U} \in $\,SU(2), which implies 
$\mbox{Det}{\mathbf U}=1$. Taking the derivative with respect to $\tau$  implies
\begin{equation}
\partial_{\tau} \mbox{Det}{\mathbf U} =0,
\end{equation}
so that
\begin{equation}
{\mathbf U}\frac{\partial}{\partial \tau} {\mathbf U}^\dagger\Big|_{1,1}=-{\mathbf U}\frac{\partial}{\partial \tau} {\mathbf U}^\dagger\Big|_{2,2}.
\end{equation}
This result was already used in obtaining Eq.~(\ref{eq:Berry}).

The next step is to construct an explicit form for $\mathbf{U}$  to obtain an explicit expression for ${\mathbf U}\frac{\partial}{\partial \tau} {\mathbf U}^\dagger\Big|_{1,1}$.
We start by considering the  rotation matrix $\mathbf R \in $\,SO(3) connecting different spin quantization axes in Eq.~(\ref{eq:action-rotinvariant}).
We choose to parametrize this rotation matrix in terms of Euler angles:
\begin{equation}
\label{Eq:Euler}
{\mathbf R}(\theta,\phi,\varphi)={\mathbf R}_0(\phi {\mathbf z} ){\mathbf R}_0(\theta {\mathbf y}) {\mathbf R}_0(\varphi {\mathbf z}),
\end{equation} 
where ${\mathbf R}_0(\gamma {\mathbf n} ) $ is a rotation around $\mathbf{n}$ by $\gamma$.
It follows from Eq.~(\ref{Eq:Euler}) that
\begin{eqnarray}
\label{eq:genRot}
\fl
{\mathbf R}(\theta,\phi,\alpha)\!\!=\!\! \Bigg ( \begin{tabular}{ c c c }
$\!\cos{\theta}\cos{\phi}\cos{\alpha}-\sin{\theta}\sin{\alpha}$ & $\!-\cos{\theta}\cos{\phi}\sin{\alpha}-\sin{\theta}\cos{\alpha}$ & $\!\cos{\theta}\sin{\phi}$ \\ 
$\!\sin{\theta}\cos{\phi}\cos{\alpha}+\cos{\theta}\sin{\alpha}$ & $\!\cos{\theta}\cos{\alpha}-\sin{\theta}\cos{\phi}\sin{\alpha}$  & $\!\sin{\theta}\sin{\phi}$ \\ 
$\!-\sin{\phi}\cos{\alpha}$ & $\!\sin{\phi}\sin{\alpha}$ & $\!\cos{\phi}$ \\ 
\end{tabular}
\Bigg ).\nonumber\\
\end{eqnarray}
Such a parametrization  is not unique and various ways of parameterizing a rotation are possible. {This issue will be addressed below.}
We will use the relation between the tensor and the spinor representation,
\begin{equation}
\label{rep}
\sum_{i}{\mathbf R}(\theta,\phi,\varphi)|_{i,3}\sigma^{i} ={\boldsymbol{\Omega}}\cdot \boldsymbol{\sigma}={\mathbf U}^{\dagger}\sigma^{3}{\mathbf U},
\end{equation}
where  $\boldsymbol{\Omega}^{\dagger}=(\cos{\theta}\sin{\phi},\sin{\theta}\sin{\phi},\cos{\phi})$  and ${\mathbf U}\,\in$\,SU(2).

A possible choice for ${\mathbf U} $ of Eq.~(\ref{rep}) is {\it e.g.} given by
\begin{eqnarray}
\label{eq:U}
{\mathbf U}(\theta(\tau),\phi(\tau))=\Bigg ( \begin{tabular}{ c c }
 $\cos{\phi(\tau)/2}$   & $e^{-i \theta(\tau)}\,\sin{\phi(\tau)/2}$ \\ 
 $-e^{i\theta(\tau)}\,\sin{\phi(\tau)/2}$ & $\cos{\phi(\tau)/2}$ \\ 
\end{tabular}
\Bigg ).
\end{eqnarray}
It should be clear from Eq.~(\ref{rep}) that $-\mathbf{U}$  is another possible choice.
With the form of Eq.~(\ref{eq:U})
it follows  that
\begin{eqnarray}
\label{eq:UdU}
{\mathbf U}\frac{\partial}{\partial \tau} {\mathbf U}^\dagger =\frac{1}{2}\Bigg ( \begin{tabular}{ c c }
 $i\,(1-\cos{\phi})\dot \theta$   & $-e^{-i\,\theta}(\dot \phi -i \dot \theta \sin\phi ) $ \\ 
 $ e^{i\,\theta}(\dot \phi +i \dot \theta \sin \phi ) $ & $ -i\,(1-\cos{\phi})\dot \theta $ \\ 
\end{tabular}
\Bigg ),
\end{eqnarray}
where $\dot \phi=\partial_\tau \phi=\partial \phi/\partial \tau$ and likewise for $\dot \theta$. 
Thus, $\int d\tau  {\mathbf U}\frac{\partial}{\partial \tau} {\mathbf U}^\dagger\Big|_{1,1}$ is indeed purely imaginary and is equal to
$i\,\frac{1}{2} \int d\tau (1-\cos{\phi(\tau)})\dot \theta(\tau)$.

The spin path integral expression of  the previous section was obtained through the parameterization of  the rotation matrix $\mathbf R\in $\,SO(3) connecting different spin quantization axes.
An alternative way to parameterize  the gauge transformation connecting different spin quantization axes can be obtained if we start by specifying the matrix $U \in \, $SU(2)
relating $\boldsymbol{\chi}$ and $\boldsymbol{\psi}$  via $\boldsymbol{\chi}^\dagger=\boldsymbol{\psi}^\dagger U$. 
 As $\sigma^3$ is a traceless 2$\times$2 matrix, $\boldsymbol{U}^\dagger\sigma^3 \boldsymbol{U}$ can be expanded in terms of $\sigma^i\,(i=1,2,3)$,
\begin{equation}
\label{eq:Hopf}
\boldsymbol{U}^\dagger\sigma^3 \boldsymbol{U}=\boldsymbol{\Omega}\cdot \boldsymbol{\sigma}.
\end{equation}
One finds $\Omega_x=\frac{\alpha+\alpha^*}{1+|\alpha|^2}$, $\Omega_y=i\frac{\alpha-\alpha^*}{1+|\alpha|^2}$, and $\Omega_z=\frac{1-|\alpha|^2}{1+|\alpha|^2}$, where we have set $\alpha=\boldsymbol{U}|_{12}/\boldsymbol{U}|_{11}$. It follows  that $\alpha\,\in \, \mathbb{C}$ and that
\begin{eqnarray}
\label{eq:UdU-version2}
{\mathbf U}\partial_\tau {\mathbf U}^\dagger =\frac{1}{2}\Bigg ( \begin{tabular}{ c c }
 $ \frac{\alpha^*\partial_\tau \alpha-\alpha \partial_\tau \alpha^*}{1+|\alpha|^2}$   & $-2\frac{\partial_\tau \alpha^*}{1+|\alpha|^2} $ \\ 
 $ 2\frac{\partial_\tau \alpha}{1+|\alpha|^2} $ & $ -\frac{\alpha^*\partial_\tau \alpha-\alpha \partial_\tau \alpha^*}{1+|\alpha|^2} $ \\ 
\end{tabular}
\Bigg ).
\end{eqnarray}
This seemingly gives rise to an alternative form of the spin path integral. The first version is based on summing {all} possible trajectories on the unit sphere, parameterized by  the two real variable $\theta$ and $\phi$. The second version is based on a single complex variable $\alpha$ and {thus is in terms of sum of all possible} trajectories in the complex plane.
Both versions are of course equivalent as should be clear from the derivation and both lead to common versions of the spin coherent states-based path integral~\cite{Perelomov,Wen,Inomata,Kuratsuji.80}.
{In fact,} the relation between $\alpha$ and the angles $\theta$ and $\phi$ is easy to obtain:
\begin{equation}
\alpha=e^{i \theta} \tan{\phi/2},
\end{equation}
which maps every point $\alpha=x+iy$ of the complex plane onto a point $\vec{\mathcal{P}}$ of the unit sphere $S^2$, $\mathcal{P}_i=\Omega_i$ ($i=x,y,z$) with $\mathcal{P}_x^2+\mathcal{P}_y^2+\mathcal{P}_z^2=1$, which is nothing but the stereographic projection.  

Above, we obtained relation Eq.~(\ref{rep}) by specifying the rotation of the quantization axis via Eq.~(\ref{eq:genRot}) and {used} the general relation between the tensor and spinor representation, {\it{e.g.}}
$\boldsymbol{R}^{lm}=\frac{1}{2} Tr \big[\sigma^l \boldsymbol{U}\sigma^m\boldsymbol{U}^\dagger\big]$.
Eq.~(\ref{rep}) or Eq.~(\ref{eq:Hopf}) define a mapping of a matrix $U\,\in $ SU(2) onto a point on the two-dimensional sphere $S^2$. The parameter space of SU(2) is the three-dimensional sphere $S^3$~\cite{Urbantke.03}. As $\boldsymbol{U}$ is not unique one may wonder how different choices affect the Berry phase term. 
We note that multiplying $\boldsymbol{U}$  by $e^{i\gamma \sigma^3}$ from the left,$\boldsymbol{U}\rightarrow e^{i\gamma \sigma^3} \boldsymbol{U}=\tilde{\boldsymbol{U}}$, leaves the right hand side of Eq.~(\ref{eq:Hopf}) invariant,
\begin{equation}
\tilde{\boldsymbol{U}}^\dagger  \sigma^3 \tilde{\boldsymbol{U}}=\boldsymbol{U}^\dagger e^{-i\gamma \sigma^3} \sigma^3 e^{i\gamma \sigma^3} \boldsymbol{U}=\boldsymbol{\Omega} \cdot \boldsymbol{\sigma},
\end{equation}
while $\alpha \rightarrow \tilde{\alpha}=\alpha e^{-2i\gamma}$. Yet, the diagonal elements of Eq.~(\ref{eq:UdU-version2}) and thus the Berry phase term are unaffected, $\int d\tau\,{\tilde{\boldsymbol{U}}}\partial_\tau {\tilde{\boldsymbol{ U}}}^\dagger\big|_{1,1}=\int d\tau \,{\mathbf U}\partial_\tau {\mathbf U}^\dagger\big|_{1,1}$, as long as $\gamma$ remains $\tau$-independent.

\subsection{Hubbard-Stratonovich decouplings}
Having discussed how different parametrizations of the gauge transformation on the local degrees of freedom give rise to different forms of the spin coherent state path integral, we now turn to a discussion of the different possible choices in decoupling the interaction term in Eq.~(\ref{eq:Anderson})  via a Hubbard-Stratonovich transformation. In Sec.~\ref{PIversion}, we choose to write 
$U n_{+}n_{-}=U/4 n^2-U S_{z}^2$ which allowed us to decouple both $n^2$ and $S_{z}^2$ in terms of two scalar Hubbard-Stratonovich fields. 
As shown, this choice leads to the well-known Kondo exchange coupling $J_K$ and potential scattering term $W$ when evaluated at the saddle point level. 
There are many alternative ways of writing the interaction term of the Anderson model in such a way that it allows for a  Hubbard-Stratonovich decoupling. We could {\it e.g.} have used 
\begin{equation}
\label{version2}
U n_{+}n_{-}=\frac{U}{2} n -\frac{2 U}{3} \boldsymbol{S}^2,
\end{equation} 
which was already used in obtaining Eq.~(\ref{eq:Anderson-v2}). $\boldsymbol{S}$ {was} defined right after Eq.~(\ref{eq:Anderson-v2}). Alternatively, we could have made use of
\begin{equation}
\label{version3}
U n_{+}n_{-}=-\frac{U}{2} n +\frac{U}{2} n^2.
\end{equation} 
While Eq.~(\ref{version3}) leads to a single scalar real Hubbard-Stratonovich field, Eq.~(\ref{version2}) gives rise to a real-valued vector decoupling field and an explicitly rotationally invariant effective action.
This then raises the question if any possible decoupling of $U n_{+}n_{-}$ could have been used in Sec.~\ref{PIversion} to obtain an effective low-energy action. The Hubbard-Stratonovich transformation is based on an identity so that any decoupling may be used as long as the functional integration over all field configurations is performed or equivalently if all fluctutations around a chosen saddle point are included.  This, however, is usually not practical.
It is easy to check that a Hubbard-Stratonovich decoupling based on either  Eq.~(\ref{version2}) or Eq.~(\ref{version3}) followed by a saddle point approximation fails to lead to the correct expression for $J_K$ and  $W$. 
The question which of the possible Hubbard-Stratonovich decouplings is the {\it right} or (rather) best one has been discussed   {\it e.g.} in Refs.~\cite{Gomes.77,Prange.81} in the context of the Hubbard model with the conclusion that the use of the operator identity $n_{\pm}^2= n_{\pm}$ in rewriting $U n_{+}n_{-}$ has to be avoided as its use induces an artifical interaction between electrons with equal spins~\cite{Gomes.77}. This was analyzed in detail by H.~Keiter, who showed how a proper summation of  diagram classes within a perturbative expansion lead to a cancellation of this artifical two-body interaction~\cite{Keiter.70}.  
Note that 
\begin{equation}
\label{version4}
U n_{+}n_{-}=\frac{U}{4} n +\frac{U}{8} n^2 -\frac{U}{2} \boldsymbol{S}^2,
\end{equation}
is yet another identity but in contrast to  Eq.~(\ref{version2}) and  Eq.~(\ref{version3}), the operator identity $n_{\pm}^2= n_{\pm}$ is not needed in its derivation. 
Nonetheless, using Eq.~(\ref{version4}) instead of Eq.~(\ref{eq:decoupling_v1}) in the derivation of Sec.~\ref{PIversion} {would} again not reproduce the correct exchange coupling $J_K$ and strength of the potential scatterer $W$. 

The origin of why Eq.~(\ref{eq:decoupling_v1}) is a reasonable choice in the present context while decouplings based on Eqs.~(\ref{version2})-(\ref{version4}) fail is easily understood.
In {\it{e.g}} Eq.~(\ref{version3}), a one-body term is separated out of the two-body interaction.  Therefore, the diagrammatic expansions of $U n_{+}n_{-}$ and $U/2 n^2$ differ by Hartree-Fock insertions.
Decoupling the  $U/2 n^2$ via a bosonic Hubbard-Stratonovic field followed by a saddle point approximation leads to a contribution of the form $\Delta_0 n$ where $\Delta_0$ is the saddle point value of the decoupling field.  This static limit  is effectively a Hartree contribution and can therefore not compensate for the extracted one-body term.  Therefore, saddle point approximations based on Hubbard-Stratonovic decouplings of Eqs.~(\ref{version2})-(\ref{version4}) will not
reproduce the Hartree-Fock approximation of the Anderson model.
That a Hubbard-Stratonovich decoupling based on Eq.~(\ref{version2}) followed by a saddle point approximation fails to reproduce the Hartree-Fock approximation of the Hubbard model was already noticed by H.~Schulz~\cite{Schulz.90}. 
In contrast, Eq.~(\ref{eq:decoupling_v1}) can be thought of as the sum of $U n_{+}n_{-}=\frac{U}{2} n -U S_z^2$ and Eq.~(\ref{version3}). Thus, the one-body terms cancel and a saddle point approximation after decoupling the squares in Eq.~(\ref{eq:decoupling_v1}) reproduces the Hartree-Fock approximation of the Anderson model.

{An additional complication that arises within the path integral formalism when performing a non-linear coordinate transformation from  Cartesian to, {\it e.g.}, polar coordinates as in Eqs.~(\ref{version2}) and (\ref{version4}), is known as the Edwards-Gulyaev effect~\cite{Edwards.64}. In the new coordinate system, terms of order higher than one in the time discretization parameter may not be negligible which  leads to correction terms. A nice introduction into the Edwards-Gulyaev effect and variable changes within a path integral can be found in the book by Inomata et al.~\cite{Inomata}}
We end this section by noting that
the local part of the Anderson model, $H_{\mbox{\tiny loc}}=\sum_{\alpha=\pm} \epsilon_d d^{\dagger}_\alpha d^{}_\alpha+Ud^{\dagger}_{+}d^{\dagger}_{-}d^{}_{-}d^{}_{+}$ can also be written as
$H_{\mbox{\tiny loc}}=2(\epsilon_d+U/2)\tau^z+(2/3)U \boldsymbol{\tau}^2+\epsilon_d$, where $\boldsymbol{\tau}=(\tau^x,\tau^y,\tau^z)$  with $\tau^x=(d^{\dagger}_{+}d^{\dagger}_{-}+d^{ }_{-}d^{ }_{+})/2$, $\tau^y=-i(d^{\dagger}_{+}d^{\dagger}_{-}-d^{ }_{-}d^{ }_{+})/2$, and $\tau^z=(\sum_{\alpha=\pm}d^{\dagger}_{\alpha}d^{}_{\alpha}-1)/2$.
At its particle-hole symmetric point, $2\epsilon_d+U=0$, $H_{\mbox{\tiny loc}}$ also possesses 
an SU(2) symmetry in the charge sector. If $U<0$ and $\epsilon_d\ll |V|^2$, spin excitations only virtually play a role  and the effective low-energy model will be a Kondo model in the charge sector giving rise to the charge Kondo effect~\cite{Taraphder.91}.
Clearly, 
repeating the steps of Sec.~\ref{PIversion} with the role of spin and charge interchanged would lead from the Anderson model to the charge Kondo model as the proper low-energy model. It would however  be interesting to
generalize the path integral version of the Schrieffer-Wolff transformation to a unified treatment of both low-energy limits such that
an 'optimal' decoupling is automatically chosen depending on the parameters of the model. This optimal decoupling has to be chosen among the possible
saddle points suitable for obtaining a low-energy action of the Anderson model as discussed above. We will return to this issue \cite{Ribeiro.16}.

\section{The magnetic transistor}
\label{Sec:magT}

The previous section established that a path integral version  of the Schrieffer-Wolff transformation applied to the Anderson model correctly yields the Kondo model with {an} anti-ferromagnetic spin-exchange coupling plus {a} potential scattering term known from the standard Schrieffer-Wolff transformation. 
In this section, we will exemplify the effectiveness of this approach by applying it to an Anderson impurity immersed in an interacting host metal $H_{\mbox{\tiny host}}$.  The presence of interactions in the host  will affect the equation of motion of $c^\dagger(\tau)_{i=0,\sigma}$, $c(\tau)_{i=0,\sigma}$, where $i=0$ marks the location of the Anderson impurity. It is thus natural to expect that the effective low-energy model of an Anderson impurity is modified by the
presence of interactions in the host metal.

For simplicity, we will describe the interaction part of the host metal Hamiltonian  by a  Hubbard term, {\it i.e.}
\begin{eqnarray}
\label{interactingLead}
H_{\mbox{\tiny host}} &=& -t  \sum_{\langle i,j\rangle,\sigma=\pm} c^{\dagger}_{i,\sigma}c^{}_{j,\sigma}+\tilde{U}\sum_{i} c^{\dagger}_{i,+}c^{\dagger}_{i,-}c^{}_{i,-}c^{}_{i,+}\\
&=& -t  \sum_{\langle i,j\rangle,\sigma=\pm} c^{\dagger}_{i,\sigma}c^{}_{j,\sigma} +\frac{\tilde{U}}{2}\sum_{i,\sigma}n_{i,\sigma} - \frac{2}{3} \tilde{U} \sum_{i} {\boldsymbol{S}}_{i} \cdot {\boldsymbol{S}}_{i}, \nonumber
\end{eqnarray} 
where $\sum_{\langle i,j\rangle}$ denotes a sum over nearest neighbors.

A perturbative  treatment of the interaction term in  Eq.(\ref{interactingLead}) within the path integral version of the Schrieffer-Wolff transformation is straightforward. 
In the following, we will however assume that the strength of the Coulomb term $ \tilde{U}$ in the host metal is sufficiently large to spontaneously break the spin-rotational invariance, {\it i.e.}, that the host is in a ferromagnetic state.
More specifically, we will consider  a quantum dot attached to ferromagnetic leads~\cite{Pasupathy.04,Kirchner.05a}.
Magnetic leads in contact with artifical nanostructures offer the possibility to utilize the spin degree of freedom to manipulating charge transport and vice versa and form a  building block for potential spintronic devices~\cite{Wolf.01}. As a result, such systems have recently attracted considerable attention. Quantum dots attached to ferromagnetic leads have been experimentally realized in a variety of systems~\cite{Pasupathy.04,Cottet.06,Krompiewski.06,Hauptmann.08}.  
These systems allow for the experimental investigation of the interplay of Kondo screening processes with magnetic excitations if the quantum dot is in the {so-called} Coulomb blockade regime. 
Such a system  {was {\it e.g.}} realized in Ref.~\cite{Pasupathy.04} {where it was}  explicitly demonstrated that complete Kondo screening can occur despite a non-vanishing spin-polarization in the leads.
The Kondo effect in a ferromagnetic host has also been investigated theoretically~\cite{Kirchner.05a,Kirchner.08b,Martinek.03,Choi.04,Krawiec.07}. 
Most of these studies do however treat the magnetism at the mean field level and thus ignore the effect of spin-wave excitations. Unlike Stoner excitations, which underlie Kondo singlet formation, spin waves are true low-energy excitations by virtue of Goldstone's theorem. As pointed out in Refs.~\cite{Kirchner.05a,Kirchner.08b}, the coupling to ferromagnetic spin waves can have a profound effect on the Kondo singlet formation and lead to the critical Kondo destruction at a quantum phase transition.

\begin{figure}
\centering{}
\includegraphics[width=0.45\linewidth]{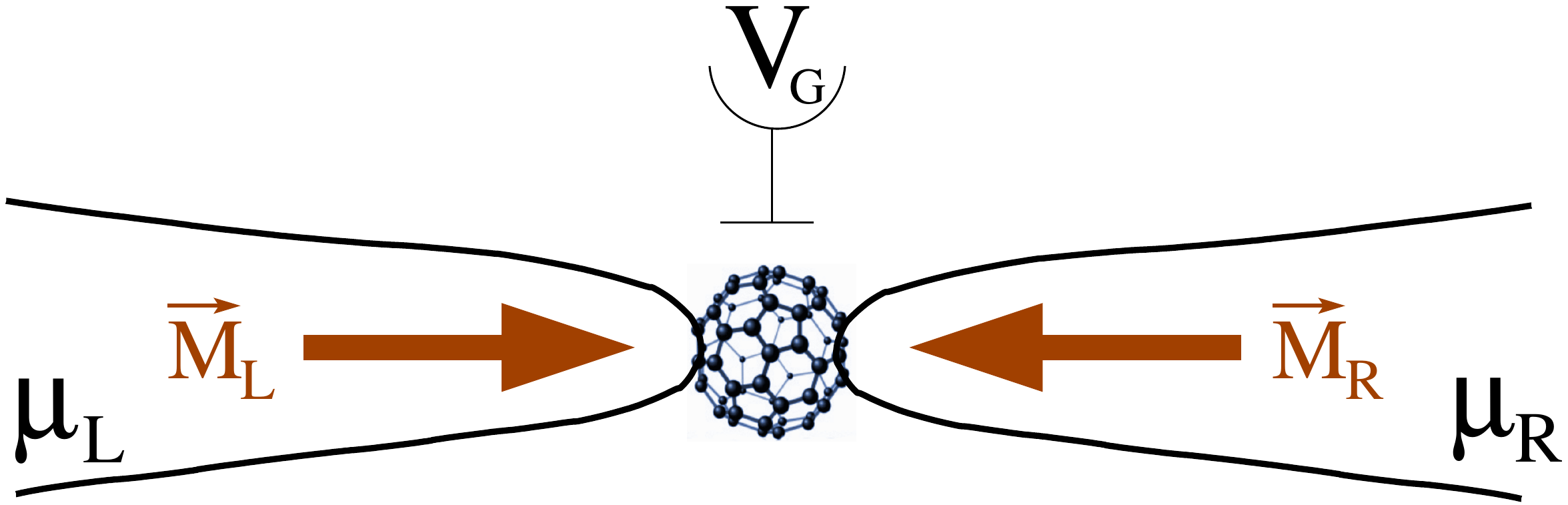}
\protect\caption{\label{figure1} Sketch of the magnetic single electron transistor of Ref.~\cite{Pasupathy.04}. A C$_{60}$ molecule that contains a Ni atom is attached to two ferromagnetic leads that act as source and drain. Applying a bias voltage $eV$ corresponds to a difference in the chemical potentials $\mu_L$ and $\mu_R$ of the two leads with $\mu_L-\mu_R=eV$. The magnetization of the leads is indicated by the red arrows and is taken to be opposite in the two leads. A gate voltage $V_G$ allows to tune the local energy levels of the molecule and thus to tune the {the ratio $g/T_K$, where $g$ is the coupling between the local moment formed in the quantum dot and the spin waves of the two leads. $T_K$ is the Kondo temperature of the system.}
}
\end{figure}

{Here,} we will demonstrate
that the coupling to the magnetic leads necessarily implies a coupling to the Goldstone bosons that accompany the breaking of the continuous spin symmetry in the leads~\cite{Kirchner.05a} and obtain the full effective low-energy model via a Schrieffer-Wolff transformation. We will find 
that the effective Hamiltonian governing the low-energy dynamics of such a structure is indeed a sub-Ohmic Bose-Fermi Kondo model with easy-plane symmetry~\cite{Zhu.02,Zarand.02} due to the $U(1)$ symmetry of the gapless spin-wave modes of the ferromagnetic leads, as argued in Ref.~\cite{Kirchner.05a}. 
The Hamiltonian of this particular sub-Ohmic Bose Fermi Kondo model is
\begin{eqnarray}
\label{eq:BFKM}
H_{\mbox{\tiny BFKM}}&=&J_K\boldsymbol{S}\cdot \boldsymbol{s}_c +\sum_{\mathbf{k},\sigma} \epsilon_{\mathbf{k}} c_{\mathbf{k},\sigma}^{\dagger}c_{\mathbf{k},\sigma}^{ }+h_{\mbox{\tiny loc}}S_z\nonumber \\
&+& g\sum_{i=x,y}\sum_{\mathbf{q}} \, S_{i}\big(\phi^{i}_{\mathbf{q},i} + \phi^{\dagger}_{-\mathbf{q},i}\big) +\sum_{i=x,y}\sum_{\mathbf{q}}\omega_{\mathbf{q}}\phi^{\dagger}_{\mathbf{q},i}\phi^{ }_{\mathbf{q},i},
\end{eqnarray}
where the spectral density of the bosons obeys $\sum_{\mathbf{q}}\delta(\omega-\omega_{\mathbf{q}})\sim \omega^{\gamma}$, with $\gamma=1/2$, $g$ is the strength of the coupling between bosons and the local spin degree of freedom and $h_{\mbox{\tiny loc}}$ is a local magnetic field. Because of the sub-Ohmic nature ($\gamma<1$) of the bosonic bath, a quantum critical point exists in the system that separates a Kondo-screened local Fermi liquid phase from a critical local moment phase~\cite{Kirchner.05a}.
There are, however, additional terms generated by the Schrieffer-Wolff transformation, as shown below.\\
We start from an Anderson model attached to two interacting leads, each described by $H_{\mbox{\tiny host}}$ of Eq.~(\ref{interactingLead}), {\it i.e.},
\begin{eqnarray}
\label{eq:QDHamiltonian}
\fl
H_{\mbox{\tiny dot}}= \sum_{\sigma=\pm} \epsilon_d d^{\dagger}_{\sigma} d^{}_{\sigma}+U d^{\dagger}_{+}d^{}_{+}d^{\dagger}_{-}d^{}_{-} +\sum_{{\bf k},\sigma=\pm,\alpha=L,R}
(V_{{\bf k},\alpha} c^{\dagger}_{{\bf k},\sigma,\alpha} d^{}_{\sigma} + \mbox{h.c.})+H_{\mbox{\tiny host}}^{\alpha},
\end{eqnarray}
where  $\alpha=L/R$ refers to the left/right lead. In what follows, it will for simplicity  be assumed that the hybridization between the quantum dot and the two leads is $\mathbf{k}$-independent ($V_{\alpha,{\mathbf k}}=V_{\alpha}$) and that the two leads are made  of the same material and have identical shapes, so that the electronic density of states (and thus the hopping $t$ in Eq.~(\ref{interactingLead}), and also $\tilde{U}$ are the same for both leads. We will also assume that the magnetization in the two leads point in opposite directions, {\it i.e.} the leads are aligned anti-parallel to each other, see Figure \ref{figure1}. \\
In the following, we will first treat the case with only one lead in detail before discussing the full problem of two anti-aligned leads.  To this end, the summation over $\alpha$ in Hamiltonian, Eq.~(\ref{eq:QDHamiltonian}), is taken to  only contain one term and the summation index will be suppressed, {\it i.e.}, $X_{\alpha}^{}\rightarrow X^{}_{}$, where $X$ is any one of the set $V_{\alpha},V^{*}_{\alpha},c^{\dagger}_{{\bf k},\sigma,\alpha}, c^{}_{{\bf k},\sigma,\alpha} $.

Following the steps that led to  Eq.~(\ref{eq:decoupledAction}) of Sec.\ref{PIversion} and with the help of a  Hubbard-Stratonovich vector decoupling field $\boldsymbol{\phi}$  to decouple the interaction term in the lead (see also \ref{AppA}),  we obtain
\begin{eqnarray}
\label{eq:MagAction}
\fl
Z=\int {\mathcal{D}}[\boldsymbol{\Omega}]\int{\mathcal{D}}[\boldsymbol{\overline{\psi}},\boldsymbol{\psi}]
\int {\mathcal{D}}[\boldsymbol{\phi},\Delta,m,\overline{\boldsymbol{\chi}},\boldsymbol{\chi}]\,
e^{-S_{\mbox{\tiny eff}}[\overline{\boldsymbol{\chi}},\boldsymbol{\chi}, \boldsymbol{\overline{\psi}},\boldsymbol{\psi},\Delta,m,\boldsymbol{\phi}]}~,\nonumber \\
\fl
S_{\mbox{\tiny eff}}[\overline{\boldsymbol{\chi}},\boldsymbol{\chi}, \boldsymbol{\overline{\psi}},\boldsymbol{\psi},\Delta,m,\boldsymbol{\phi}] =
\int_0^\beta d\tau \Bigg\{  \overline{\boldsymbol{\chi}} (\partial_{\tau}+\epsilon_d+U^\dagger \partial_{\tau}U) \boldsymbol{\chi}+\frac{1}{U}(\Delta^2+m^2) \nonumber \\
\fl
-\frac{1}{2}\sum_{\mathbf{k}} \boldsymbol{\phi}^{}({\mathbf{k}})\cdot\boldsymbol{\phi}^{}(-\mathbf{k})
+ \sum_{{\mathbf{k}},{\mathbf{k}'}} \boldsymbol{\overline{\psi}}_{}({\mathbf{k}'})\Big ( \big (\partial_{\tau}+\epsilon_{\mathbf{k}}^{ } -\mu \big)\delta({\mathbf{k}}-{\mathbf{k}'}) + \sqrt{\frac{\tilde{U}}{3}}  \boldsymbol{\phi}^{ }(\mathbf{k}-\mathbf{k}') \boldsymbol{\sigma} \Big ) \boldsymbol{\psi}_{ }({\mathbf{k}})
\Big . \nonumber \\
\fl
- i\Delta \overline{\boldsymbol{\chi}}\boldsymbol{{\chi}}-m \overline{\boldsymbol{\chi}}\sigma^z_{}\boldsymbol{{\chi}}
+\sum_{{\mathbf{k}}}\big [ V^{}_{ } \overline{\boldsymbol{\chi}} \boldsymbol{\psi}_{ }({\mathbf{k}}) +V^{*}_{ } \boldsymbol{\overline{\psi}}_{ }({\mathbf{k}})\boldsymbol{\chi}\big ] 
 \Big .  \Bigg \}.
\end{eqnarray}
The action is of the form $S=S^{\mbox{\tiny dot}}+S^{\mbox{\tiny lead}}_{ }+S^{\mbox{\tiny dot-lead }}_{ }$, where $S^{\mbox{\tiny dot}}$ is the quantum dot action independent of the lead, $S^{\mbox{\tiny lead}}_{}$, is the action associated with the  lead, and $S^{\mbox{\tiny dot-lead }}_{}$  describes the coupling between the lead and the quantum dot.

When integrating out the local degree of freedom, the coupling term in the action between the local  $\overline{\boldsymbol{\chi}} (\boldsymbol{\chi})$ and  the conduction electron fields $\overline{\boldsymbol{\psi}} (\boldsymbol{\psi})$ generates a term that  possesses the following form
\begin{eqnarray}
\label{eq:hybrid}
\fl
\sum_{\mathbf{k},\mathbf{k'}} V^{*}_{}V^{}_{}\overline{\boldsymbol{\psi}}_{ }(\mathbf{k}) {\mathbf U}^{\dagger} G_d {\mathbf U} 
\boldsymbol{\psi}_{ }(\mathbf{k'})=
|V|^2 \overline{\boldsymbol{\psi}}_{ }(0) {\mathbf U}^{\dagger} G_d {\mathbf U} 
\boldsymbol{\psi}_{ }(0)\,=\, \overline{\boldsymbol{\psi}}_{ }(0) \Big (\frac{J_K}{2} \boldsymbol{\Omega}\cdot  \frac{\boldsymbol{\sigma}}{2} + W  \Big ) \boldsymbol{\psi}_{ }(0),
\end{eqnarray}
where, in the last step, the saddle point {values} for the local decoupling fields $\Delta$ and $m$,{\it i.e.}, $\Delta_0=iU/2$ and $m_0=U/2$ were used, see Eq.~(\ref{eq:couplings}).
As before, this amounts to
 $J_K=\frac{2|V|^2 U}{|\epsilon_d|(\epsilon_d+U}$, $W=|V|^2\frac{\epsilon_d+U/2}{|\epsilon_d|(\epsilon_d+U)}$.
Constant terms in the action have been absorbed into the measure of the functional integral.
We also introduced $\boldsymbol{\psi}_{}(0)=\sum_{\mathbf{k}} \boldsymbol{\psi}_{ }(\mathbf{k})$, which is the field value of $\boldsymbol{\psi}_{ }({\mathbf{r=0}})$  at the location of the quantum dot and likewise for  $\overline{\boldsymbol{\psi}}_{}(0)$.
The unitary matrix  ${\mathbf U}$ in Eq.~(\ref{eq:hybrid}) relates the local Grassmann fields and is defined in Sec.~\ref{PIversion} between Eqs.~(\ref{eq:action-rotinvariant}) and (\ref{eq:decoupledAction}).\\
So far, the treatment parallels the one of Sec.~\ref{PIversion} {despite} of the presence of interactions in the lead. The next step is to take the saddle point value of the local (in configuration space) decoupling field $\boldsymbol{\phi}$ and consider the Gaussian fluctuations around this saddle point. As described in detail in \ref{AppA}, a local gauge transformation on the conduction electron fields generates the spin-wave action. This local gauge transformation on the conduction electron fields 
will modify  the coupling term of Eq.~(\ref{eq:hybrid}). With
$\overline{\boldsymbol{\psi}}_{ }(\tau)=\overline{\boldsymbol{\xi}}_{ }(\tau){\mathbf V}(\tau)$ and $\boldsymbol{\psi}_{ }(\tau)={\mathbf V}^\dagger(\tau)\boldsymbol{\xi}_{ }(\tau)$ we obtain
\begin{equation}
\overline{\boldsymbol{\psi}}_{ }(0) \Big (\frac{J_K}{2} \boldsymbol{\Omega}\cdot  \frac{\boldsymbol{\sigma}}{2} + W  \Big ) \boldsymbol{\psi}_{ }(0)=\overline{\boldsymbol{\xi}}_{ }(0) {\mathbf V} \Big (\frac{J_K}{2} \boldsymbol{\Omega}\cdot  \frac{\boldsymbol{\sigma}}{2} + W  \Big ) {\mathbf V}^\dagger\boldsymbol{\xi}_{ }(0),
\end{equation}
where 
the transformation matrix ${\mathbf V}^{\dagger}$   is given by
\begin{eqnarray}
\label{eq:TrafoV}
{\mathbf V}^{\dagger}\, =\,\Bigg ( \begin{tabular}{ c c }
 $1-\frac{1}{2} \bar{\alpha}\alpha $ & $ -\alpha $ \\ 
 $ \bar{\alpha} $ & $ 1-\frac{1}{2} \bar{\alpha}\alpha $ \\ 
\end{tabular}
\Bigg ),
\end{eqnarray}
as shown in  \ref{AppA}.
Here, $\alpha=(\delta \phi_x-i\delta \phi_y)/(2\phi_0)$ where $\delta \phi_x$ and $\delta \phi_y$ are
 Gaussian fluctuations around the saddle point value $\boldsymbol{\phi}_0$   and are perpendicular to the direction of the magnetization. The saddle point value of $\boldsymbol{\phi}_0$ is related to the magnetization in the lead via $\boldsymbol{\phi}_0=-\sqrt{\tilde{U}}/3\langle \mathbf{S}\rangle$. 
The next step is to use Eq.(\ref{rep}), $\sum_{i} {\mathbf R}^{ij} \sigma^{i}={\mathbf V}\sigma^{j} {\mathbf V}^{\dagger}$, where the elements of the rotation matrix  ${\mathbf R}$
can be obtained from
\begin{equation}
\label{eq:Tr}
 {\mathbf R}^{lm}=\frac{1}{2}\mbox{Tr}\Big\{ \sigma^{l}{\mathbf V}\sigma^{m} {\mathbf V}^{\dagger} \Big \}.
\end{equation}
The explicit form of $\mathbf{R}$ suggests to introduce the matrix ${\mathbf{\tilde{R}}}$ via ${\mathbf R}=\mathbf{I}+2\mathbf{\tilde{R}}$, where $\mathbf{I}$ represents the three-dimensional unit matrix.
One finds for ${\mathbf{\tilde{R}}}$ up to quadratic order in $\alpha~(\bar{\alpha})$, {\it i.e.}, within the spin-wave approximation,
\begin{eqnarray}
\label{eq:matrixR2}
{\mathbf{\tilde{R}}}= \Bigg ( \begin{tabular}{ c c c }
$\re{\alpha}\re{\alpha}$~ & $-\im{\alpha}\re{\alpha}$~ & $\re{\alpha}$ \\ 
$-\im{\alpha}\re{\alpha}$~ & $\re{\alpha}\re{\alpha}$~   & $\im{\alpha}$ \\ 
$-\re{\alpha}$ & $-\im{\alpha}$ & $-\bar{\alpha}\alpha$ \\ 
\end{tabular}
\Bigg ).
\end{eqnarray}
This implies 
\begin{eqnarray}
\label{eq:hybrid3}
\fl
\frac{J_K}{2} \overline{\boldsymbol{\xi}}_{ }(0)  \boldsymbol{\Omega}\cdot {\mathbf V}^{}_{} \frac{\boldsymbol{\sigma}}{2}{\mathbf V}^\dagger_{}\boldsymbol{\xi}_{ }(0)  +\overline{\boldsymbol{\xi}}_{ }(0) W \boldsymbol{\xi}_{ }(0)\,=\, \overline{\boldsymbol{\xi}}_{ }(0)   \Big (\frac{J_K}{2} \boldsymbol{\Omega}\cdot  \frac{\boldsymbol{\sigma}}{2} + W + \frac{J_K}{2}  \boldsymbol{\Omega}\cdot {\mathbf{\tilde{R}}} \boldsymbol{\sigma} \Big)\boldsymbol{\xi}_{ }(0).
\end{eqnarray}
Thus, we arrive at a form of the effective action of a quantum dot attached to a magnetic lead 
\begin{eqnarray}
\label{eq:MagAction2}
\fl
Z=\int {\mathcal{D}}[\boldsymbol{\Omega}]\int{\mathcal{D}}[\overline{\boldsymbol{\xi}},\boldsymbol{\xi}]
\int {\mathcal{D}}[\overline{\alpha},\alpha]\,
e^{-S_{\mbox{\tiny eff}}[ \overline{\boldsymbol{\xi}},\boldsymbol{\xi},\overline{\alpha},\alpha,\boldsymbol{\Omega}]}~,\nonumber \\
\fl
S_{\mbox{\tiny eff}}[ \boldsymbol{\overline{\psi}},\boldsymbol{\psi},\overline{\alpha},\alpha,\boldsymbol{\Omega}] =
\int_0^\beta d\tau  {\mathbf U}\frac{\partial}{\partial \tau} {\mathbf U}^\dagger\Big|_{1,1}
\,+\,\int_0^\beta d\tau \Big\{  
\sum_{\mathbf{q}} \overline{\alpha}(\mathbf{q})\big( \partial_\tau -\omega_{\mathbf{q}}\big )\alpha(\mathbf{q})
 \nonumber \\
\fl
+ \sum_{{\mathbf{k}}} \overline{\boldsymbol{\xi}}_{}({\mathbf{k}}) G_c^{-1} \boldsymbol{\xi}_{}({\mathbf{k}}) + \overline{\boldsymbol{\xi}}_{ }(0)   \Big ( \frac{J_K}{2} \boldsymbol{\Omega}\cdot  \frac{\boldsymbol{\sigma}}{2} + W + \frac{J_K}{2}  \boldsymbol{\Omega}\cdot {\mathbf{\tilde{R}}} \boldsymbol{\sigma} \Big)\boldsymbol{\xi}_{ }(0)
 \Big .  \Big \},
\end{eqnarray}
where  $G_c^{-1}=\partial_{\tau}+\epsilon_{\mathbf{k}} -\mu+\phi_0\sqrt{\tilde{U}/3}\sqrt{1+(\delta \phi_x/\phi_0)^2+(\delta \phi_y/\phi_0)^2}\sigma^3$. 
As the action in Eq.~(\ref{eq:MagAction2}) is local in imaginary time, it corresponds to an effective low-energy Hamiltonian. So far, no assumptions other than
the spin-wave approximation and those underlying the derivation of the results of  Sec.~\ref{PIversion}, which are warranted in the local moment regime, have  been made.

In the following, the effective action will be cast into a more convenient form. 
Using Eq.~(\ref{eq:matrixR2}), we have
\begin{eqnarray}
\label{eq:MagAction3}
Z=\int {\mathcal{D}}[\boldsymbol{\Omega}]\int{\mathcal{D}}[\overline{\boldsymbol{\xi}},\boldsymbol{\xi}]
\int {\mathcal{D}}[\overline{\alpha},\alpha]\,
e^{-S_{\mbox{\tiny eff}}[ \boldsymbol{\overline{\xi}},\boldsymbol{\xi},\overline{\alpha},\alpha,\boldsymbol{\Omega}]}~,\nonumber \\
\fl
S_{\mbox{\tiny eff}}[ \boldsymbol{\overline{\psi}},\boldsymbol{\psi},\overline{\alpha},\alpha,\boldsymbol{\Omega}] =
\int_0^\beta d\tau  {\mathbf U}\frac{\partial}{\partial \tau} {\mathbf U}^\dagger\Big|_{1,1}
+\int_0^\beta d\tau \Bigg\{  
\sum_{\mathbf{q}} \overline{\alpha}(\mathbf{q})\big( \partial_\tau -\omega_{\mathbf{q}}\big )\alpha(\mathbf{q})
\nonumber \\
\fl
+ \sum_{{\mathbf{k}}} \boldsymbol{\overline{\xi}}_{}({\mathbf{k}}) G_c^{-1} \boldsymbol{\xi}_{}({\mathbf{k}}) +  \frac{J_K}{2} \,
\big ( \begin{tabular}{ c c  }
$\Omega_x$~ & $\Omega_y$ \\  
\end{tabular}
\big ) \cdot
 \Bigg ( \begin{tabular}{ c   }
$s_c^x$\\ $s_c^y$ \\  
\end{tabular}
\Bigg )
-\frac{3}{2} \sqrt{\rho_0} J_K \Omega_z \phi_0
\nonumber \\
\fl
+ \sum_{\mathbf{k},\mathbf{k'}} W \overline{\boldsymbol{\xi}}_{ }(\mathbf{k}) \boldsymbol{\xi}_{ }(\mathbf{k'}) 
-3 J_K \sqrt{\rho_0}\phi_0 \,
\big ( \begin{tabular}{ c c  }
$\Omega_x$~ & $\Omega_y$ \\  
\end{tabular}
\big ) \cdot
 \Bigg ( \begin{tabular}{ c   }
$\re{\alpha(\mathbf{r}=0)}$\\ $\im{\alpha(\mathbf{r}=0)}$ \\  
\end{tabular}
\Bigg )
\nonumber \\
\fl
+  J_K\,\big ( \begin{tabular}{ c c  }
$\Omega_x$~ & $\Omega_y$ \\  
\end{tabular}
\big )
 \Bigg ( \begin{tabular}{ c c  }
$\re{\alpha}\re{\alpha}$~ & $-\im{\alpha}\re{\alpha}$ \\ 
$-\im{\alpha}\re{\alpha}$~ & $\re{\alpha}\re{\alpha}$ \\  
\end{tabular}
\Bigg )
 \Bigg ( \begin{tabular}{ c   }
$s^x_c$\\ $s^y_c$ \\  
\end{tabular}
\Bigg )\nonumber \\
\fl
+\, J_K\,\Omega_z\big(-\re{\alpha}s_c^x-\im{\alpha}s_c^y+3 \sqrt{\rho_0} \phi_0\overline{\alpha}\alpha\big) 
 \Big .  \Bigg \},
\end{eqnarray}
where $\langle {s}^{z}_c \rangle=-3/\sqrt{\tilde{U}}\phi_0$ and $\rho_0 \tilde{U}\gtrsim 1$, 
{{\it i.e.},} the Stoner criterion for itinerant ferromagnets (in the form $\rho_0 \tilde{U}= 1$), {was} used. In this expression, $\rho_0$ represents the conduction electron density of states at the Fermi level and $s_c^i$ is the $i$-th component of the electron spin density at the quantum dot. 
Instead of expressing the spin-wave excitations through a complex bosonic field $\alpha=\re{\alpha}+i\im{\alpha}$ it will be more convenient to introduce two real bosonic fields, $\phi^x=\re{\alpha}=\bar{\phi}^x$ and $\phi^y=\im{\alpha}=\bar{\phi}^y$. As these fields are real, we have for their Fourier transforms $\phi^{x/y}_{\mathbf{q}}\equiv \phi^{x/y}(\mathbf{q})=\phi^{x/y}(-\mathbf{q})\equiv \phi^{x/y}_{\mathbf{-q}}$. 
Therefore,
\begin{eqnarray}
\label{eq:MagAction4}
\fl
Z=\int {\mathcal{D}}[\boldsymbol{\Omega}]\int{\mathcal{D}}[\overline{\boldsymbol{\xi}},\boldsymbol{\xi}]
\int {\mathcal{D}}[\bar{\phi}^x,\phi^x,\bar{\phi}^y,\phi^y]\,
e^{-S_{\mbox{\tiny eff}}[ \overline{\boldsymbol{\xi}},\boldsymbol{\xi},\bar{\phi}^x,\phi^x,\bar{\phi}^y,\phi^y,\boldsymbol{\Omega}]}~, 
\\
\fl
S_{\mbox{\tiny eff}}[ \overline{\boldsymbol{\xi}},\boldsymbol{\xi},\bar{\phi}^x,\phi^x,\bar{\phi}^y,\phi^y,\boldsymbol{\Omega}] =
\int_0^\beta d\tau  {\mathbf U}\frac{\partial}{\partial \tau} {\mathbf U}^\dagger\Big|_{1,1} 
\,+\,\int_0^\beta d\tau \Bigg\{  
\sum_{i=x,y} \sum_{\mathbf{q}} \bar{\phi}^{i}_{\mathbf{q}}\big( \partial_\tau -\omega_{\mathbf{q}}\big )\phi^{i}_{\mathbf{q}}
\nonumber \\
\fl
+ \sum_{{\mathbf{k}}} \overline{\boldsymbol{\xi}}_{}({\mathbf{k}}) G_c^{-1} \boldsymbol{\xi}_{}({\mathbf{k}})  +  \frac{J_K}{2} \,
\big ( \begin{tabular}{ c c  }
$\Omega_x$~ & $\Omega_y$ \\  
\end{tabular}
\big ) \cdot
 \Bigg ( \begin{tabular}{ c   }
$s_c^x$\\ $s_c^y$ \\  
\end{tabular}
\Bigg )
-\frac{3}{2} \sqrt{\rho_0} J_K \Omega_z \phi_0
\nonumber \\
\fl
+ \sum_{\mathbf{k},\mathbf{k'}} W \overline{\boldsymbol{\xi}}_{ }(\mathbf{k}) \boldsymbol{\xi}_{ }(\mathbf{k'}) 
-3 J_K \sqrt{\rho_0}\phi_0 \,\sum_{\mathbf{q}}
\big ( \begin{tabular}{ c c  }
$\Omega_x$~ & $\Omega_y$ \\  
\end{tabular}
\big ) \cdot
 \Bigg ( \begin{tabular}{ c   }
$\bar{\phi}^x_{\mathbf{q}}+\phi^x_{-\mathbf{q}}$\\ $\bar{\phi}^y_{\mathbf{q}}+\phi^y_{-\mathbf{q}}$ \\  
\end{tabular}
\Bigg )
 \Big .  \Bigg \}
+K_1+K_2 +K_3, \nonumber
\end{eqnarray}
where the terms $K_1$, $K_2$, and $K_3$ are given by
\numparts
\begin{eqnarray}
\fl
\label{K1K3a}
K_{1} =  J_K \int_0^\beta d\tau \,\big ( \begin{tabular}{ c c  }
$\Omega_x$~ & $\Omega_y$ \\  
\end{tabular}
\big )
\Bigg ( \begin{tabular}{ c c  }
$\bar{\phi}^x(0)\phi^x(0)$~ & $-\bar{\phi}^x(0)\phi^y(0)$ \\ 
$-\bar{\phi}^y(0)\phi^x(0)$~ & $\bar{\phi}^x(0)\phi^x(0)$ \\  
\end{tabular}
\Bigg )
 \Bigg ( \begin{tabular}{ c   }
$s^x_c$\\ $s^y_c$ \\  
\end{tabular}
\Bigg ),
\\
\fl
\label{K1K3b}
K_{2}= -J_K \sum_{\mathbf{q}}\int_0^\beta d\tau \, \Omega_z \sum_{\mathbf{q}}\big( (\bar{\phi}^x_{\mathbf{q}}+\phi^x_{-\mathbf{q}}) s_c^x+(\bar{\phi}^y_{\mathbf{q}}+\phi^y_{-\mathbf{q}}) s_c^y \big ),
\\
\fl
\label{K1K3c}
K_{3}=3J_K \sqrt{\rho_0}\phi_0 \int_0^\beta d\tau \, \Omega_z (\bar{\phi}^x(0)\phi^x(0)+\bar{\phi}^y(0)\phi^y(0)).
\end{eqnarray}
\endnumparts
Thus, it can be seen that the effective action is that of a general Bose-Fermi Kondo model with additional coupling terms that lead to $K_1$, $K_2$, and $K_3$ in the effective action. The coupling constant $g$ between the quantum spin and the bosonic bath in Eq.~(\ref{eq:BFKM}) is given by $g=-3J_K\sqrt{\rho_0}\phi_0$. The sign of g is irrelevant and can {\it e.g.} be removed using the XY-symmetry of the model.

The effect of the additional terms that are present in the low-energy model of the magnetic single-electron transistor but  are not part of the Hamiltonian $H_{\mbox{\tiny BFKM}}$ of Eq.~({\ref{eq:BFKM}}), {\it i.e.} the terms $K1$, $K_2$, and $K_3$ in Eqs.~(\ref{K1K3a})-(\ref{K1K3c}), will be analyzed in the following using scaling arguments. As a result, these terms will turn out to be irrelevant.
The tree-level scaling dimension will be obtained with respect to $J_K=0$ and $g=0$~\cite{Zhu.02}.
It was pointed out in Refs.\cite{Kirchner.05a,Kirchner.08b} that ferromagnetic spin-waves can give rise to a sub-Ohmic bosonic bath due {to} their quadratic dispersion, {\it i.e.} $\int d^3 {\mathbf{q}} \delta(\omega-\omega_{\mathbf{q}}) \sim \omega^{\gamma}$ with $\gamma=1/2$. Therefore, $\langle\phi_{i}({\mathbf{r}}=0,\tau) \phi_{i}^\dagger({\mathbf{r}}=0,0)\rangle\sim 1/\tau^{1+\gamma}$ ($i=x,y$) for large $\tau$ and $\phi_{i}$ scales as $\tau^{-(1+\gamma)/2}$. Similarly, one finds that $\boldsymbol{s}_c({\mathbf{r}}=0)$ scales as $\tau^{-1}$ and that $\boldsymbol{S}$ scales as $\tau^{0}$, where $\boldsymbol{S}$ is the local spin operator of  Eq.~({\ref{eq:BFKM}}) that gives rise to the field $\boldsymbol{\Omega}$ in Eq.~(\ref{eq:MagAction4}). For the engineering dimensions of the coupling constants in Eq.~({\ref{eq:BFKM}}) we thus find $[J_K]=0$, where $[\hat{x}]$ denotes the engineering dimension of $\hat{x}$. For the coupling to the bosonic bath we find $[g]=(1-\gamma)/2$ which implies that $g$ is relevant for $\gamma<1$. The additional couplings in Eq.~(\ref{K1K3a}) have the form
\begin{equation}
g_{1}^{ijk} \int_0^\beta d\tau \, \Omega_{i}(\tau) \bar{\phi}^{j}_{ }(\mathbf{r}=0,\tau)\phi^{j}_{ }(\mathbf{r}=0,\tau)s_{c}^{k}(\tau),
\end{equation}
with $i,j,k \in {x,y}$ possessing engineering dimensions $[g_{1}^{ijk}]=1+\gamma$ and thus are irrelevant for all $\gamma>-1$. For 
\begin{equation}
g_{2}\sum_{i=x,y}\sum_{\mathbf{q}}\int_0^\beta d\tau \, \Omega_{z}(\tau) (\bar{\phi}^{i}_{\mathbf{q}}(\tau)+\phi^{i}_{-\mathbf{q}}(\tau))s_{c}^{i}(\tau)
\end{equation}
one finds that $[g_{2}]=(1+\gamma)/2$, so that this term also remains irrelevant as long as $\gamma>-1$. For $g_3$ we find from
\begin{equation}
g_{3} \sum_{i=x,y}\int_0^\beta d\tau \, \Omega_{z}(\tau) \bar{\phi}^{i}_{ }(\mathbf{r}=0,\tau)\phi^{i}_{ }(\mathbf{r}=0,\tau)
\end{equation}
that $[g_{3}]=\gamma$ and that this coupling is irrelevant for all $\gamma>0$. In the present case we have $\gamma=1/2$ and thus the term $K_1$, $K_2$, and $K_3$ in the effective action of Eq.~(\ref{eq:MagAction4}) are irrelevant and can be ignored in the low-energy limit.

So far, the case of  a quantum impurity immersed in a ferromagnetic metal has been considered, where the magnetization leads to a spin polarization or net magnetic field along the direction of $\boldsymbol{\phi}$ at the impurity site. 
In the ferromagnetic transistor, two leads, {\it i.e.} the source and drain lead, are taken to have opposite spin polarization so that the effective polarization cancels at the site of the quantum dot. In the following we will generalize the derivation to this case~\footnote{A generalization of our results to an arbitrary angle between the magnetization in the two leads is straightforward.}. The corresponding Hamiltonian $H_{\mbox{\tiny BFKM}}$ was already introduced in Eq.~(\ref{eq:QDHamiltonian}), at the beginning of this section. We will label the leads as in Fig.~\ref{figure1}, so that in what follows the index $L$ ($R$) refers to the left (right) lead. 
If the spin-polarization in the leads vanishes, a simple mapping exists to an effective lead
with effective hybridization $V_{\mbox{\tiny eff}}=\sqrt{|V_L|^2+|V_R|^2}$.
{In that case, we have}
%
%
%
\begin{eqnarray}
\label{eq:diagon}
\sum_{\alpha,\beta=L,R} \bar{\boldsymbol{\psi}}_{\alpha}V^{*}_{\alpha}{\mathbf U}^\dagger G_d {\mathbf U}V^{}_{\beta} \boldsymbol{\psi}_{\beta}=
\big ( \begin{tabular}{ c c  }
$\overline{\boldsymbol{\psi}}_{L}$~ & $\overline{\boldsymbol{\psi}}_{R}$ \\  
\end{tabular}
\big )
\boldsymbol{A}
 \Bigg ( \begin{tabular}{ c   }
$\boldsymbol{\psi}_{L}$\\ $\boldsymbol{\psi}_{R}$ \\  
\end{tabular}
\Bigg )\nonumber \\
= \big ( \begin{tabular}{ c c  }
$\overline{\boldsymbol{\psi}}_{A}$~ & $\overline{\boldsymbol{\psi}}_{S}$ \\  
\end{tabular}
\big ) \boldsymbol{\Delta}_{\mathbf A}
 \Bigg ( \begin{tabular}{ c   }
$\boldsymbol{\psi}_{A}$\\ $\boldsymbol{\psi}_{S}$ \\  
\end{tabular}
\Bigg ),
\end{eqnarray}
where
the coupling matrix $\mathbf{A}$ has eigenvalues $0$ and $V_{\mbox{\tiny eff}}^2{\mathbf U}^\dagger G_d {\mathbf U} $ and the matrix $\mathbf{M}$ is chosen such that $\boldsymbol{\Delta}_{\mathbf A}={\mathbf M_{}^{-1}}
 {\mathbf{A}} {\mathbf M}$ is diagonal, {\it i.e.},
  \begin{eqnarray}
{\mathbf M}=\frac{1}{V_{\mbox{\tiny eff}}}
 \left ( \begin{tabular}{ c c }
$-V_R^{*}$~ & $V_L^{}$~ \\ 
$V_L^{*}$~  &$V_R^{}$~  \\ 
\end{tabular}
\right ).
\end{eqnarray}

The transformation to the new basis therefore is given by $\big ( \begin{tabular}{ c c  }
$\overline{\boldsymbol{\psi}}_{A}$~ & $\overline{\boldsymbol{\psi}}_{S}$ \\  
\end{tabular}
\big )=\big ( \begin{tabular}{ c c  }
$\overline{\boldsymbol{\psi}}_{L}$~ & $\overline{\boldsymbol{\psi}}_{R}$ \\  
\end{tabular}\big ){\mathbf{M}}$. The quadratic terms of the conduction electrons in the new {basis}  remain diagonal in the label $S$ and $A$ if $\epsilon_{\mathbf k}^{L}=\epsilon_{\mathbf k}^{R}$. As a result, only the symmetric  combination of the lead  states will couple to the quantum dot.\\
In the case of finite spin-polarization in the leads, the transformation is slightly  more involved. 
Our assumptions of identical leads and oppositely aligned magnetization amount to $\epsilon_{\mathbf k}^L=\epsilon_{\mathbf k}^R$ and $\boldsymbol{\phi}^R_0=-\boldsymbol{\phi}^L_0$. The change in magnetization from left to right lead can be described by a $\pi$-rotation: $\boldsymbol{\phi}^R ={\mathbf R}(\boldsymbol{n},\pi) \boldsymbol{\phi}^L$
The rotation matrix ${\mathbf R}$ of a $\pi$-rotation around $\boldsymbol{n}=\pm(\sin \gamma ,\cos \gamma ,0)$  is
\begin{eqnarray}
\label{eq:piRot}
{\mathbf R}(\boldsymbol{n},\pi)=\Bigg ( \begin{tabular}{ c c c }
$-\cos \gamma $~ & $\sin \gamma $~ & $0$ \\ 
$\sin \gamma $~ & $\cos \gamma $~   & $0$ \\ 
$0$ & $0$ & $-1$ \\ 
\end{tabular}
\Bigg ).
\end{eqnarray}
By virtue of the $XY$-symmetry perpendicular to $\boldsymbol{\phi}_0$ of the action of  Eq.~(\ref{eq:MagAction}) all  $\boldsymbol{n}$ turn out to be equivalent.
Additionally, note that one has the freedom to change  
$\boldsymbol{\phi}^{}=\phi_{0}^{}\vec{e}_z+\delta \phi_x^{}\vec{e}_x +\delta \phi_y^{}\vec{e}_y$ to {\it e.g.}
$\boldsymbol{\phi}^{}=\phi_{0}^{}\vec{e}_z-\delta \phi_x^{}\vec{e}_x +\delta \phi_y^{}\vec{e}_y$
(see \ref{AppA}).
This change will only affect the off-diagonal elements of Eq.~(\ref{eq:TrafoV}) (or Eq.~(\ref{eq:trafoV}) of \ref{AppA}) but leaves Eq.~(\ref{appA7}) invariant. The relation between the magnetization of the two leads becomes $\boldsymbol{\phi}^R_0=-\boldsymbol{\phi}^L_0$, $\re{\alpha^R(\mathbf{r}=0)}=-\re{\alpha^L(\mathbf{r}=0)}$, and $\im{\alpha^R(\mathbf{r}=0)}=-\im{\alpha^L(\mathbf{r}=0)}$\footnote{Alternatively, the transformation matrix $\boldsymbol{V}_{R}^{\dagger}$ of Eq.~(\ref{eq:TrafoV}) for the right lead can be related to the one for the left lead by $\boldsymbol{V}_R^{\dagger}=i \sigma^2 \boldsymbol{V}_{L}^{\dagger}(i\sigma^2)^{-1}_{}$.}.
The next step is to generalize the transformations of Eq.~(\ref{eq:diagon}) to the 4-component spinor 
\big ( \begin{tabular}{ c c c c }
$\overline{\psi}_{L}^{\uparrow}$~ & $\overline{\psi}_{L}^{\downarrow}$~& $ \overline{\psi}_{R}^{\uparrow}$~ & $\overline{\psi}_{R}^{\downarrow}$\\  
\end{tabular}
\big ).
The hybridization matrix remains singular and possesses eigenvalues $0,0,|V_L|^2+|V_R|^2,|V_L|^2+|V_R|^2$ so that the transformation to the symmetric/anti-symmetric basis outlined above can be carried through. 
The coupling between the local spin and the spin waves in the two leads add up. For identical couplings to left and right leads, {\it i.e.} $V_L=V_R$, the terms $K_1$ and $K_2$ and the local magnetic field at the dot site vanishes completely. For the Kondo coupling, we find that the prefactor of $\Omega_z s^z_c$ vanishes. Nonetheless, we can add the term $\frac{J_K}{2} \Omega_z s^z_c$ to the effective action as the flow towards the strong-coupling fixed point restores SU(2) symmetry.

As a result, we have shown that the effective low-energy model of a quantum dot attached to ferromagnetic leads with anti-aligned magnetization is that of a sub-Ohmic Bose-Ferm Kondo model whose Hamiltonian is given by
\begin{eqnarray}
H_{\mbox{\tiny BFKM}}=J_{K} \sum_{k,k',\sigma,\sigma'} \boldsymbol{S}\cdot c^{\dagger}_{k,\sigma}\frac{\boldsymbol{\tau}}{2}c^{}_{k',\sigma'}
+\sum_{k,\sigma}\epsilon_k c^{\dagger}_{k,\sigma} c^{ }_{k,\sigma} +\sum_{\sigma,\sigma'}\sum_{\boldsymbol{k},\boldsymbol{k'}} W c_{\boldsymbol{k},\sigma}^{\dagger}c_{\boldsymbol{k'},\sigma'}^{}
\nonumber \\
+ g\sum_{i=x,y}\sum_{\mathbf{q}} \, S_{i}\big(a^{i}_{\mathbf{q},i} + a^{\dagger}_{-\mathbf{q},i}\big) + h_{\mbox{\tiny loc}} S_z
+\sum_{q} \omega_q \boldsymbol{a}^{\dagger}_{q}\cdot \boldsymbol{a}^{}_{q}. \nonumber
\end{eqnarray}
For the effective Kondo coupling $J_K$, the coupling to the bosonic bath $g$ and the strength of the potential scattering term, one finds
\numparts
\begin{eqnarray}
J_K&=&2 V_{\mbox{\tiny eff}}^2 \frac{U}{|\epsilon_d|(\epsilon_d+U)},\\
\label{g-coupling}
g&=&6 V_{\mbox{\tiny eff}}^2 \frac{U\sqrt{\rho_0}\phi_0}{|\epsilon_d|(\epsilon_d+U)},\\
h_{\mbox{\tiny loc}}&=& 3(|V|_R^2-|V|_L^2)\sqrt{\rho_0}U/(|\epsilon_d|(\epsilon_d+U)),\\
W&=& V_{\mbox{\tiny eff}}^2 \frac{\epsilon_d+U/2}{|\epsilon_d|(\epsilon_d+U)},
\end{eqnarray}
\endnumparts
where $|V_L|$ and $|V_R|$ denote the hybridization strength between the quantum dot and the left and right leads, $\epsilon_d$ and $U$ are local energies of the quantum dot, and $\rho_0$ is the density of states of the leads at the dot size and $\phi_0$ is proportional to the lead magnetization. 
The spectral density of the bosons, $\sum_{\boldsymbol{q}}\delta(\omega-\omega_{\boldsymbol{q}})\sim \omega^\gamma$,  is sub-Ohmic ($\gamma<1$) as a consequence of the quadratic dispersion of ferromagnetic spin waves. Explicitly, one finds  $\gamma=1/2$ so that the phase diagram of this model contains a zero-temperature phase transition that separates a Kondo-screened from a Kondo-desroyed phase~\cite{Kirchner.05a}. Tuning the gate voltage ($V_G$ in Fig.~\ref{figure1}) changes the charging energy {$\Delta$ of the quantum dot. As a result the tuning parameter of the (sub-Ohmic) Bose-Fermi Kondo-model,
$g/T_K$, where $T_K\sim  \exp[1/(\rho_0 J_K)]/\rho_0$, varies over a huge range which allows to tune the system across the quantum phase transition~\cite{Kirchner.05a}.}
This makes the magnetic transistor an ideal system to experimentally explore Kondo-destroying quantum criticality in and out of equilibrium.
{How different levels of approximation affect the parameter dependence of the coupling $g$ is an interesting question.}
{
Regardless, whether the coupling $g$ is linear in $1/\Delta$, as appearing here, or quadratic in $1/\Delta$, as appearing in Ref.~\cite{Kirchner.05a}, it is able to induce a Kondo-destruction quantum critical point through a competition with the Kondo energy scale, which depends on $\Delta$ exponentially.}

\section{Conclusion}
\label{Sec:conclusion}

We have shown how to perform the Schrieffer-Wolff transformation within the path integral formulation. The Schrieffer-Wolff transformation projects the dynamics of the original Hamiltonian $H$ into a subspace of the Hilbert space associated with $H$ corresponding to the low-energy sector of the model. These steps are conveniently carried out at the level of the action associated with $H$. An effective action that is local in (imaginary) time can be related back to an effective Hamiltonian. In this process it is the dynamic phase which is determined by the geometry of the state space accessible to the system and which in turn determines the quantum nature of the effective degree of freedom.  
In the case discussed here the low-energy sector of the Hilbert space of the local degree of freedom was identified with the coset space SU(2)/U(1)$=S^2$. Thus, the effective low-energy model is that of a quantum spin.
The path integral version of the Schrieffer-Wolff transformation not only brings out the topological features associated with the resulting quantum spin model but also leads to
simplifications associated with  those of the path integral formalism over the operator formalism. This should be particularly helpful when a knowledge of  higher order corrections is required or in deriving the effective low-energy model for more complex systems than the standard single-impurity Anderson model. 
As an explicit demonstration of the advantages of the path integral version over the operator form of the Schrieffer-Wolff transformation, we derived the effective low-energy model of a quantum dot attached to two interacting leads with spontaneously broken spin-rotational invariance. Future applications should include  multi-impurity Anderson models~\cite{Ong.11} and multi-level quantum dots attached to superconducting and magnetic leads. 
We also analyzed in which way different decouplings and saddle points affect the final result and discussed the charge analog of the spin Kondo model. 
Our work thus adds new insights to a classic problem of strong correlation physics. 
In deriving our results, we started from the finite-$U$ Anderson model, Eq.~(\ref{eq:Anderson}). Nonetheless, the limit $U\rightarrow \infty$ can be considered in $J_{k,k'}$ and $W_{k.k'}$ of the resulting model, Eq.~(\ref{eq:KondoHam}). 
How a Schrieffer-Wolff transformation  can be performed if the limit $U\rightarrow \infty$ is taken from the beginning, or, equivalently, a pseudoparticle representation of the Hamiltonian is {employed}, is an interesting question which deserves further attention~\cite{Tuengler.95,Fresard.01}.\\

\noindent
{\textbf{Acknowledgments:}}
We thank Petr Jizba, Johann Kroha, Lawrence S.~Schulmann, Qimiao~Si and V\'itor Vieira for useful discussions.
P.~Ribeiro acknowledges support by FCT through the Investigador FCT contract IF/00347/2014.
S.~Kirchner acknowledges partial support by the National Science Foundation of China, grant No.11474250.
This work was in part performed at the Aspen Center for Physics, which is supported by
National Science Foundation grant PHY-1066293.

\appendix
\section{Dynamics of Spin Waves}
\label{AppA}
\setcounter{section}{1}

We review in the following the derivation of the effective action of an itinerant ferromagnet  described in terms of an one-band Hubbard-like model. Although this is largely textbook material, see {\it e.g.} chapter 3 of Ref.~\cite{Fradkin}, our reason for doing so is two-fold. Firstly, we show that the spin wave dynamics is the result of a Berry phase term which can be obtained in a manner similar to that in our presentation of the Schrieffer-Wolff transformation. For a general treatment of Goldstone boson dynamics in terms of generalized coherent states we refer to the work by M.~Blasone and P.~Jizba\cite{Blasone.12}. 
Secondly, the summary of the spin wave dynamics presented here is needed in the application of the Schrieffer-Wolff transformation for the magnetic  transistor given in Sec.~\ref{Sec:magT}.

We start with the Hubbard model, {\it i.e.}
\begin{eqnarray}
 H_{\mbox{\tiny }}&=& -t  \sum_{\langle i,j\rangle,\sigma=\pm} c^{\dagger}_{i,\sigma}c^{}_{j,\sigma}+\tilde{U}\sum_{i} c^{\dagger}_{i,+}c^{\dagger}_{i,-}c^{}_{i,-}c^{}_{i,+}\\
&=& -t  \sum_{\langle i,j\rangle,\sigma=\pm} c^{\dagger}_{i,\sigma}c^{}_{j,\sigma} +\frac{\tilde{U}}{2}\sum_{i,\sigma}n_{i,\sigma} - \frac{2}{3} \tilde{U} \sum_{i} {\boldsymbol{S}}_{i} \cdot {\boldsymbol{S}}_{i}, \nonumber
\end{eqnarray} 
where $\sum_{\langle i,j\rangle}$ denotes a sum over nearest neighbors.

 We will decouple the interaction part of  $H_{\mbox{\tiny }} $ in terms of a real Hubbard-Stratonovich vector decoupling field $\boldsymbol{\phi}_{}$, using
\begin{equation}
\int d \boldsymbol{\phi}_{i} \exp [-\frac{1}{2} (\boldsymbol{\phi}_{i})^2 +\sqrt{\tilde{U}/3}\boldsymbol{\phi}_{i}\cdot \boldsymbol{S}_i^{}]=(2\pi)^{3/2}\exp [\frac{\tilde{U}}{6} \boldsymbol{S}_i^{}\cdot \boldsymbol{S}_i^{}].
\end{equation}
Thus, one obtains
\begin{eqnarray}
\label{appA3}
Z=\int{\mathcal{D}}[\boldsymbol{\bar{\psi}},\boldsymbol{\psi}]
\int {\mathcal{D}}[\boldsymbol{\phi}^{}]\,
e^{-S_{\mbox{\tiny eff}}[\boldsymbol{\bar{\psi}},\boldsymbol{\psi},\boldsymbol{\phi}^{}]}~,\nonumber \\
S_{\mbox{\tiny eff}}[ \boldsymbol{\bar{\psi}},\boldsymbol{\psi},\boldsymbol{\phi}^{}] =
\int_0^\beta d\tau \Bigg\{ -\frac{1}{2}\sum_{\mathbf{k}} \boldsymbol{\phi}^{}({\mathbf{k}})\cdot\boldsymbol{\phi}^{}(-\mathbf{k})\nonumber \\ 
+ \sum_{{\mathbf{k}},{\mathbf{k}'}} \boldsymbol{\bar{\psi}}_{}({\mathbf{k}'})\Big ( \big (\partial_{\tau}+\epsilon_{\mathbf{k}} -\mu \big)\delta({\mathbf{k}}-{\mathbf{k}'}) + \sqrt{\frac{\tilde{U}}{3}}  \boldsymbol{\phi}^{}(\mathbf{k}-\mathbf{k}') \boldsymbol{\sigma} \Big ) \boldsymbol{\psi}_{}({\mathbf{k}})
\Big .
 \Big .  \Bigg \}.
\end{eqnarray}
Being a Hubbard-Stratonovich decoupling field, $\boldsymbol{\phi}^{}$ in Eq.~(\ref{appA3}) does not possess a dynamical (or Berry) phase term $\bar{\boldsymbol{\phi}}^{}\partial_{\tau}\boldsymbol{\phi}^{}$.
We will assume that the spin-rotational invariance of the leads has been spontaneously broken and that the saddle point action plus Gaussian fluctuations give a proper description of the electronic and magnetic excitation spectrum of the magnetic leads of Sec.~\ref{Sec:magT}.
The saddle point value $\boldsymbol{\phi}^{}_{0}$ of the vector field $\boldsymbol{\phi}^{}$ follows from
\begin{equation}
\label{saddlepoint}
\frac{\partial \ln Z^{\mbox{\tiny }}_{}}{\partial \phi^{}_i}\Bigg |_{\phi^{}_i=\phi^{}_{i,0}}\stackrel{!}{=}0,
\end{equation}
where $\phi^{}_i$ is the $i$th component of vector $\boldsymbol{\phi}^{}$ which is related to the magnetization  through $\boldsymbol{\phi}^{}(\mathbf{r}) =-2\sqrt{\tilde{U}}/{3}\langle \boldsymbol{S}(\mathbf{r})\rangle$.
For the ferromagnetic case considered here, $\boldsymbol{\phi}_{0}$ is spatially constant: $\boldsymbol{\phi}_{0}(k-k')=\boldsymbol{\phi}_{0}\delta(k-k')$.
We choose the magnetization to be along the $\hat{z}$-direction. 
The Gaussian fluctuations $\delta \phi_x^{}, \delta \phi_y^{} \perp \boldsymbol{\phi}_{0}$ around the saddle point solution $\boldsymbol{\phi}_{0}$ describe ferromagnetic spin waves.  They are gapless, possess $XY$-symmetry, and have  a quadratic dispersion. We set $\boldsymbol{\phi}^{}=\phi_{0}^{}\vec{e}_z+\delta \phi_x^{}\vec{e}_x +\delta \phi_y^{}\vec{e}_y$. 
The fluctuations render $G_c$ ($G_{c}^{-1}=\partial_{\tau}+\epsilon_k-\mu+|\phi_0|\sqrt{\tilde{U}/3}\sigma^3-\Sigma$) non-diagonal in the basis where $\boldsymbol{\phi}_{0}=\phi_{0}^{}\vec{e}_z$. 
As the $\boldsymbol{\bar{\psi}}_{}\boldsymbol{\phi}^{}\boldsymbol{\psi}_{}$ term is local in configuration space, a local $\tau$-dependent gauge transformation can be performed: 
\begin{equation}
\label{eq:gT}
\boldsymbol{\bar{\chi}}_{i}=\boldsymbol{\bar{\psi}}_{i} {\mathbf V}^{\dagger}_{i}(\tau),~~~ \boldsymbol{\chi}_{i}={\mathbf V}^{ }_{i}(\tau)\boldsymbol{\psi}_{i}, 
\end{equation}
 such that the spin quantization axis is always along $\boldsymbol{\phi}$, {\it i.e.},
\begin{equation}
\label{spinwavegauge}
\boldsymbol{\Omega}\cdot \boldsymbol{\sigma}={\mathbf V}^\dagger \sigma^3 {\mathbf V},
\end{equation}
with $\boldsymbol{\Omega}=\boldsymbol{\phi}/|\boldsymbol{\phi}|$.
Thus, Eq.~(\ref{appA3}) becomes
\begin{eqnarray}
\label{appA6}
Z=\int {\mathcal{D}}[\delta{\phi}^{}_x,\delta \phi_y^{}]
\int{\mathcal{D}}[\boldsymbol{\bar{\chi}},\boldsymbol{\chi}]\,
e^{-S_{\mbox{\tiny eff}}[\boldsymbol{\bar{\chi}},\boldsymbol{\chi},\boldsymbol{\phi}^{}]}~,\nonumber \\
 S_{\mbox{\tiny eff}}[ \boldsymbol{\bar{\chi}},\boldsymbol{\chi},\boldsymbol{\phi}^{}] =
\int_0^\beta d\tau \Bigg\{ -\frac{1}{2}\sum_{\mathbf{k}} \boldsymbol{\phi}^{}({\mathbf{k}})\cdot\boldsymbol{\phi}^{}(-\mathbf{k})
+ \sum_{{\mathbf{k}}} \boldsymbol{\bar{\chi}}_{}({\mathbf{k}}) G_c^{-1} \boldsymbol{\chi}_{}({\mathbf{k}}) \nonumber \\
- \sum_{\langle i,j \rangle} \boldsymbol{\bar{\chi}}_{i}^{} \Big(  {\mathbf V}^{ }_{} \partial_{\tau}{\mathbf V}^{\dagger}_{} \delta_{i,j} +{\mathbf \Delta}_{i,j}^{}\Big ) \boldsymbol{\chi}_{j}^{}
\Big .
 \Big .  \Bigg \},
\end{eqnarray}
where $G_c^{-1}=\partial_{\tau}+\epsilon_{\mathbf{k}} -\mu+|\phi_0|\sqrt{\tilde{U}/3}\sqrt{1+(\delta \phi_x/\phi_0)^2+(\delta \phi_y/\phi_0)^2}\sigma^3$ and \mbox{${\mathbf \Delta}_{i,j}={\mathbf V}^\dagger_{i}({\mathbf V}^{}_{j}-{\mathbf V}^{}_{i})$}. In the continuum limit, ${\mathbf \Delta}_{i,j}\longrightarrow {\mathbf V}^{\dagger}_{} \nabla {\mathbf V}^{}_{}$.

Thus,
\begin{eqnarray}
\label{eq:perturbed}
Z=\int {\mathcal{D}}[\delta{\phi}^{}_x,\delta \phi_y^{}] \exp \Big[ -\frac{1}{2} \int_0^\beta d\tau \sum_{\mathbf{q}} \boldsymbol{\phi}^{}({\mathbf{q}})\cdot\boldsymbol{\phi}^{}(-\mathbf{q})\Big ] \\
\times
\int{\mathcal{D}}[\boldsymbol{\bar{\chi}},\boldsymbol{\chi}]\,\exp \Big[-\int_0^\beta d\tau \sum_{{\mathbf{k}}} \boldsymbol{\bar{\chi}}_{}({\mathbf{k}}) G_c^{-1} \boldsymbol{\chi}_{}({\mathbf{k}})\Big]\nonumber \\
\times
\int{\mathcal{D}}[\boldsymbol{\bar{\chi}},\boldsymbol{\chi}]\,\exp \Big[-\int_0^\beta d\tau \sum_{{\mathbf{k}}} \boldsymbol{\bar{\chi}}_{}({\mathbf{k}})G_c^{-1} \boldsymbol{\chi}_{}({\mathbf{k}}) \Big] \nonumber \\
\fl
\exp \Big[\int_0^\beta d\tau \sum_{\langle i,j \rangle} \boldsymbol{\bar{\chi}}_{i}\Big( {\mathbf V}^{ }_{}\partial_{\tau}{\mathbf V}^{\dagger}_{}  \delta_{i,j}+{\mathbf \Delta}_{i,j}   \Big)\boldsymbol{\chi}_{j}\Big]\Bigg/
\int{\mathcal{D}}[\boldsymbol{\bar{\chi}},\boldsymbol{\chi}]\,\exp \Big[-\int_0^\beta d\tau \sum_{{\mathbf{k}}} \boldsymbol{\bar{\chi}}_{}({\mathbf{k}}) G_c^{-1} \boldsymbol{\chi}_{}({\mathbf{k}})\Big], \nonumber
\end{eqnarray}
and, to linear order in ${\mathbf V}^{}_{}\partial_{\tau}{\mathbf V}^{\dagger}_{}$, we find that
$\langle  e^{\int_0^\beta d\tau \boldsymbol{\bar{\chi}}_{} V^{ }_{}\partial_{\tau}V^{\dagger}_{}\boldsymbol{\chi}_{}} \rangle \stackrel{.}{=}\exp[\int_0^\beta d\tau {\mathbf V}^{}_{}\partial_{\tau}{\mathbf V}^{\dagger}_{}\big|_{1,1}]$, see also Eq.~(\ref{eq:Berry}).

In the vicinity of the saddle point, $\delta \phi_i\ll \phi$.  Neglecting terms higher than quadratic in $\delta \phi$, one finds
\begin{eqnarray}
\label{eq:gaugetrans}
{\mathbf V}^{}_{ }\, =\,\Bigg ( \begin{tabular}{ c c }
 $1-\frac{1}{8} \big(\frac{\delta \phi_x}{\phi_0} \big)^2-\frac{1}{8} \big(\frac{\delta \phi_y}{\phi_0} \big)^2$ & $ \frac{\delta \phi_x}{2|\phi_0|} -i \frac{\delta \phi_y}{2|\phi_0|}$ \\ 
 $ -\frac{\delta \phi_x}{2|\phi_0|}  -i \frac{\delta \phi_y}{2|\phi_0|} $ & $ 1-\frac{1}{8} \big(\frac{\delta \phi_x}{\phi_0} \big)^2-\frac{1}{8} \big(\frac{\delta \phi_y}{\phi_0} \big)^2 $ \\ 
\end{tabular}
\Bigg ).
\end{eqnarray}
Eq.(\ref{eq:gaugetrans}) implies
 \begin{equation}
 {\mathbf V}^{}_{}\partial_{\tau}{\mathbf V}^{\dagger}_{}\big|_{1,1}=\frac{1}{2}\frac{\bar{\alpha}\partial_\tau \alpha-\alpha \partial_\tau \bar{\alpha}}{1+|\alpha|^2},
 \end{equation}
where $\alpha=(\delta\phi_x-i\delta\phi_y)/(2|\phi_0|)$. At the level of the Gaussian approximation we thus have $\int_0^\beta d\tau {\mathbf V}^{ }_{}\partial_{\tau}{\mathbf V}^{\dagger}_{}\big|_{1,1}=\int_0^\beta d\tau \bar{\alpha}\partial_\tau \alpha$ for each mode (after partial integration and use of periodic boundary conditions). This is the dynamical phase term associated with  a bosonic field $\alpha$. This identification allows us to express the effective action in terms of $\alpha$ and the Grassmann fields $\bar{\chi}, \chi$. 

With a change of the integration variables
and up to an overall prefactor from the saddle point value of the action, we have 
\begin{eqnarray}
\label{appA7}
\fl
Z=\int {\mathcal{D}}[\bar{\alpha},\alpha]
\int{\mathcal{D}}[\boldsymbol{\bar{\chi}},\boldsymbol{\chi}]\,
e^{-S_{\mbox{\tiny eff}}[\boldsymbol{\bar{\chi}},\boldsymbol{\chi},\boldsymbol{\phi}^{}]}~, \\
\fl
S_{\mbox{\tiny eff}}[ \boldsymbol{\bar{\chi}},\boldsymbol{\chi},\boldsymbol{\phi}^{}] =
\int_0^\beta d\tau \Bigg\{ \sum_{\mathbf{q}} \bar{\alpha}(\mathbf{q})\partial_\tau \alpha(\mathbf{q})-\sum_{\mathbf{q}} \omega_{\mathbf{q}}\bar{\alpha}^{}({\mathbf{q}})\alpha^{}(-\mathbf{q})
+ \sum_{{\mathbf{k}}} \boldsymbol{\bar{\chi}}_{}({\mathbf{k}}) G_c^{-1} \boldsymbol{\chi}_{}({\mathbf{k}})
\Big .
 \Big .  \Bigg \}. \nonumber
\end{eqnarray} 
 Terms linear in the deviation from the saddle point vanish by virtue of Eq.~(\ref{saddlepoint}).  At linear order,  $\Delta_{i,j}$ does not contribute to Eq.~(\ref{eq:perturbed}) but the next order term determines the dispersion $\omega_{\mathbf q}$ of the field $\alpha({\mathbf{q}})$, which has the property $\omega_{\mathbf q} \sim \mathbf{q}^2$ for small $|\mathbf{q}|$.\\
The transformation matrix $V$ of Eq.~(\ref{eq:gaugetrans}) in terms of the fields $\alpha$, $\bar{\alpha}$ assumes the simple form
\begin{eqnarray}
\label{eq:trafoV}
{\mathbf V}\, =\,\Bigg ( \begin{tabular}{ c c }
 $1-\frac{1}{2} \bar{\alpha}\alpha $ & $ \alpha $ \\ 
 $ -\bar{\alpha} $ & $ 1-\frac{1}{2} \bar{\alpha}\alpha $ \\ 
\end{tabular}
\Bigg ).
\end{eqnarray}

With the help of the local gauge transformation of Eq.~(\ref{eq:gT}),  Eq.~(\ref{appA3}) is equivalent to Eq.~(\ref{appA7}) within the Gaussian approximation, {\it i.e.} up to terms quadratic in $\alpha(\bar{\alpha})$. A result that has been used  in the derivation of the effective low-energy model of a quantum dot attached to ferromagnetic leads in Sec.~\ref{Sec:magT}.

\end{document}